\documentclass[twocolumn,nodate,showpacs,preprintnumbers,nofootinbib,amsmath,amssymb,aps,prd]{revtex4}

\usepackage{epsf,epsfig,graphics,graphicx}
\usepackage{verbatim,color,ulem}
\usepackage{setspace}
\usepackage{graphicx,wrapfig}
\usepackage{enumerate}
\usepackage{color}
\usepackage[ddmmyyyy]{datetime}
\usepackage{mathrsfs}
\usepackage{subfigure}
\usepackage{amssymb,amsmath,amsthm}
\usepackage[mathscr]{eucal}

\newcommand{\be}{\begin{equation}}
\newcommand{\ee}{\end{equation}}
\newcommand{\bea}{\begin{eqnarray}}
\newcommand{\eea}{\end{eqnarray}}
\newcommand{\ba}{\begin{array}}
\newcommand{\ea}{\end{array}}

\newcommand{\ra} {\rho_{10}}
\newcommand{\rb} {\rho_{20}}

\newcommand{\la} {\lambda_{11}}
\newcommand{\lb} {\lambda_{22}}
\newcommand{\lab} {\lambda_{12}}
\newcommand{\bl} {\bar\lambda}
\newcommand{\no} {\nonumber\\}

\newcommand{\ob} {\omega_{\chi,k}}
\newcommand{\ca} {c_{\phi}}
\newcommand{\cb} {c_{\chi}}

\def\be{\nopagebreak[3]\begin{equation}}
\def\ee{\end{equation}}
\def\ba{\nopagebreak[3]\begin{eqnarray}}
\def\ea{\end{eqnarray}}

\begin{document}

\title{Analogue stochastic gravity phenomena in two-component Bose-Einstein condensates: Sound cone fluctuations}
\date{\today}
\author{Wei-Can Syu}
\email{syuweican@gmail.com}
\affiliation{Department of Physics,
National Dong-Hwa University, Hualien, Taiwan, Republic of China}
\author{Da-Shin Lee}
\email{dslee@gms.ndhu.edu.tw} \affiliation{Department of Physics,
National Dong-Hwa University, Hualien, Taiwan, Republic of China}
\author{Chi-Yong Lin}
\email{lcyong@gms.ndhu.edu.tw} \affiliation{Department of
Physics, National Dong-Hwa University, Hualien, Taiwan, Republic of China}
\begin{abstract}
We investigate the properties of the condensates of cold atoms at zero temperature in the tunable binary Bose-Einstein condensate system  with a Rabi transition between  atomic hyperfine states.
We use this system to examine the effect of quantum fluctuations in a tunable quantum
gas on phonon propagation. We show that the system
can be represented by a coupled two-field model of a gapless  phonon and a
gapped  mode, which are analogous to the Goldstone and Higgs particles in particle physics. We then further  trace out the gapped modes to
give an effective purely phononic theory using closed-time-path formalism. In particular, we are interested in the sound cone
fluctuations due to
the  variation of the speed-of-sound acoustic metric, induced
by quantum fluctuations of the gapped modes.
These fluctuations can
be interpreted as inducing a stochastic space-time, and thus are regarded as  analogue phenomena of light cone
fluctuations presumably arising from quantum gravity
effects. The effects of
fluctuations can be displayed in the variation in the travel time of sound waves. We suggest
%introduce
the relevant experiments to discuss  the possibility of experimental observations.
\end{abstract}

\maketitle
\section{Introduction}
Analog models of gravity are the attempts to use laboratory systems for modeling various phenomena of general relativity (e.g., black holes or cosmological geometries), which require a deeper understanding
of (stochastic) semiclassical gravity and the role of the even more
inaccessible Planck scale physics.
Specifically,
 the pioneering work of Unruh~\cite{unruh}  was the use of sound waves in a moving fluid as an analogue for light waves in a curved spacetime, and Unruh showed  that supersonic fluid can generate a ``dumb hole", the acoustic analogue of a ``black hole", to theoretically demonstrate the existence of phononic Hawking radiation from the acoustic horizon. The program of ``analogue
gravity" to explore these analogies has received much attention and interest among particle physicists, astrophysicists and condensed matter
theorists
%since its development
(for a review, see~\cite{bar}).

In semiclassical gravity, matter is represented by quantum matter fields that propagate according to the theory of quantum fields in curved spacetime. The backreaction of quantum matter fields to the spacetime is through the semiclassical Einstein equation with sources given by the expectation value of the quantum field stress-energy tensor. However, their quantum fluctuations  are
expected to generate
%also inducing
the fluctuations of spacetime geometries.
 To explore the induced fluctuating spacetime geometries, the stochastic semiclassical gravity  is developed based upon the Einstein-Langevin equations, which involve additional sources due to the vacuum expectation value of the quantum field stress-energy bitensor~\cite{hu}.
Nevertheless, these stochastic spacetimes can arise  also from the quantum nature of the gravitational field, namely  gravitons.
One particularly striking effect of spacetime geometry fluctuations
is that a fixed light cone does not exist, giving rise to light cone fluctuations with their tiny effects on  the blurring of the images of astronomical objects, or luminosity fluctuations~\cite{ford1,ford2,ford3,ford}.
By analogy,  gapless modes (phonons) propagate causally according to  the metrics of sound cones.  However, the intrinsically quantum mechanical nature of  environment degrees of freedom which couple to the phonons may  make the cones ``fuzzy". As a result, this may induce fluctuations in their times of flight (TOF).
Some of the related effects for photons~\cite{shi,lor} and phonons~\cite{gur,kre}
have been studied by introducing phenomenological random media.

In the last several years, one of the great achievements  has been the construction of condensates in cold atom systems, in which the scattering length between atoms and therefore the speed  of sound are tunable via Feshbach resonance. Systems whose causal behavior is so simply controlled are potentially good candidates with which to explore analogies. In particular, the authors of ~\cite{dslee1,dslee2} considered
condensates of ultracold Fermi gases, which show a range of behavior, interpolating from a BCS regime characterized by Cooper pairs to a Bose-Einstein condensate (BEC) regime characterized by diatoms or dimers.
The system can be represented by a coupled two-field model of a gapless  phonon and a gapped  mode.
The gapped diatom density
fluctuations with  the known dynamics in this system
give the sound-cone
metric a stochastic component. Thus,
zero temperature
quantum fluctuations disturb the sound-cone
metric, giving fluctuations in the TOF, which provides an interesting analogue model to mimic light cone fluctuations induced by quantum gravitational effects. However, the TOF fluctuations in cold fermion systems are small, just about large enough to be measured in the current technologies.
Here we plan to work on binary BEC systems, which involve more tunable experimental parameters and larger TOF fluctuation effects are expected.
%Here we plan to work on binary BEC systems with more tunable experimental  parameters to look for more larger TOF %fluctuations effect.

With the rapid experimental progress in controlling quantum gases,
multicomponent atomic gases nowadays have been an active research field in cold atom systems.  In particular, two-component BECs  with a variety of dynamic behaviors have been experimentally studied
using the mixture of atoms with two hyperfine states of $^{87}\text{Rb}$~\cite{exp_same_atoms} or the mixture of two different species of atoms~\cite{exp_diff_atoms}.  In the ``analogue gravity" program,   the class of two-component BECs subject to a laser- or radio-waves-induced Rabi transition between different atomic hyperfine states  has been proposed to serve as an ``emergent" spacetime model, which provides  very rich spacetime geometries, such as a specific
class of pseudo-Finsler geometries, and
both bimetric pseudo-Riemannian geometries and single metric pseudo-Riemannian geometries of
interest in cosmology and general relativity~\cite{visser,wein}.
In fact, this class of the two-component BEC systems with the Rabi interaction exhibits two types of  excitations on condensates: one phonon mode due to the ``in-phase" oscillations between  two respective density waves of the binary system and
a gapped mode stemming from the ``out-of-phase" oscillations of the density waves in the presence of the Rabi transition.
In this work, we plan to consider this type of the binary BEC system. The idea is to treat the phonons as the subsystem of interest, and the gapped modes as the environment to be coarse-grained later with their effects on the dynamics of the phonons. We will adopt the closed-time-path formalism to work on the coarse-graining process with environmental effects encoded in the so-called influence functional. Subsequently, the Langevin equation of the phonons under the semiclassical approximation is derived with the noise terms given by quantum fluctuations of the gapped modes so as to determine the sound cone, along which the sound waves propagate.
The calculable fluctuations of the gapped modes provide the random medium with which the phonons scatter, giving sound cone fluctuations.
From these, the quantum fluctuations in phonon times of flight can be determined. Roughly, they can be as large as  about a 15 \% effect on the propagation time of waves across a typical condensate to be potentially observable.

Our paper is organized as follows. In the next section, we first introduce the model of the binary BEC system and analyze
 the various mean-field ground states of homogeneous condensates by tuning the parameters of the system.
In Sec.~\ref{sec2}, we consider quantum fluctuations on the mean-field ground states, which end up with two coupled modes. Then, by constructing an explicit form of the Bogoliubov transformation, the dispersion relations of the photons and the gapped modes are obtained  with their respective propagation speed as well as the effective mass of the gapped mode.
In Sec.~\ref{sec3}, we highlight the
idea of open quantum systems and
introduce the method of the closed-time-path formalism.
Starting from the whole system of  system-plus-environment,  the
environmental degrees of freedom of the gapped modes in the full density matrix are
traced over to obtain the reduced density matrix of the system of the phonons.
Their effects are all encoded in the influence functional, incorporating with
nonequilibrium correlators constructed by the gapped modes. Then,
the stochastic effective action is constructed by introducing the noise terms where the extremum of the action gives the Langevin equation that governs the propagation of the sound waves with the stochastic sound cone metric.
In Sec.~\ref{sec4}, we discuss the experiments of relevance to our study on the measurement of the flight time of the sound waves across the condensates and, with realistic experimental parameters, the flight time variance is estimated due to the stochasticity of the sound cone.
Concluding remarks are in Sec.~\ref{sec5}.

\section{The model and background condensates} \label{sec1}
Here we start from considering the binary BECs in the atomic hyperfine states of two different species of atoms, and denote them by a label 1 or 2. The action of the system in $D$-spatial dimensions  reads~\cite{tim}
\begin{align}
S=&\int{dt \, d^D x\,\mathcal{L}}\no
=& \int dt \, d^Dx \Bigg{\{} i\Big{(}\psi^*_1\frac{\partial\psi_1}{\partial t}+\psi^*_2\frac{\partial \psi_2}{\partial t}\Big{)}\no
&\quad- \sum_{j=1}^2\Big{[}\frac{1}{2m_j}|\nabla\psi_j|^2+(-1)^j\frac{\delta}{2}|\psi_j|^2+\frac{g_{jj}}{2}|\psi_j|^4\Big{]}\no
&\quad-g_{12}|\psi_1|^2|\psi_2|^2-\frac{\Omega}{2}(\psi_1\psi_2^*+\psi_1^*\psi_2)\Bigg{\}}\, ,
\label{eq:1}
\end{align}
where  $\psi_j (x)$ is the field operator of the atom in the hyperfine
$j$ state with atomic mass $m_j$ for $j=1,2$.
The ultracold atomic collisions can be effectively described by a contact
pseudopotential parametrized only by the scattering length. Consequently the strength
of the atomic interaction is
proportional to the scattering length with
the coupling constant $g_{jj}$ for  the intraspecies interaction within each $j$ state, and the coupling constant $g_{12}$ for the interspecies interaction between $j=1$ and $j=2$ states.
The relation between the coupling constant $g$ and the scattering length $a$ depends on the spatial dimensionality $D$ to be specified later.
Moreover, the reduced mass is $m_{12}=m_1m_2/(m_1+m_2)$. The scattering length can be tuned via a homogeneous magnetic field by the use of the Feshbach resonances.
The binary system can also be considered as
the same atoms in two different internal states  by
simply taking the equal mass limit, namely $m_1=m_2=m$.
In addition, we introduce a Rabi or Josephson coupling term, which can be  introduced by a laser field or radio waves, causing a conversion between two hyperfine states with the effects proportional to the Rabi-frequency $\Omega$~\cite{tim,lin03}.  Applying the laser field can also effectively shift the energy level of the internal states characterized by the detuning $\delta$~\cite{tim}.
In this model, $\Omega$ and $\delta$ are tunable parameters.

The quantum field $\psi_j(x)$ and its Hermitian
conjugate satisfy equal-time commutation relations
\begin{eqnarray}
&& [\psi_i ({\bf x}, t), \psi^\dagger_j ({\bf x}',t)]  = \delta_{ij} \, \delta^{(D)}({\bf x}- {\bf x}'),\nonumber\\
&& [\psi_i ({\bf x}, t),\psi_j ({\bf x}',t)] =[\psi_i^\dagger({\bf x}, t),\psi_j^\dagger({\bf x}',t)]=0\, .
\end{eqnarray}
The Lagrangian $\mathcal{L}$ is invariant under the $U(1)$ gauge
transformation
\bea
\psi_j ({\bf x}, t) && \rightarrow  e^{i\theta}\psi_j({\bf x}, t),\nonumber \\
\psi^\dagger_j ({\bf x}, t) && \rightarrow e^{-i\theta}\psi^\dagger_j ({\bf x}, t) \, ,
\eea
where $\theta$ is a constant phase. The consequence of this $U(1)$
gauge symmetry is conservation of the number of atoms.
Thus, the number operator of atoms below,
\begin{equation}
N=N_1+N_2= \int d^D x \,\sum_{j=1}^2 \psi^\dagger_j ({\bf x}, t)\psi_j ({\bf x}, t) \,
\label{n1_n2}
\end{equation}
is a conserved quantity.
In a Bose-condensed gas, the condensate plays a crucial
role and hence two condensate wave functions, which are expectation values of the
respective field operators $\langle \psi_j ({\bf x},t) \rangle$, are described by their densities and phases
as
\begin{equation}
\langle \psi_j ({\bf x},t) \rangle =\sqrt{\rho_{j0}} \, e^{i \theta_{j0}}  ,\label{mean_f_decompose}
\end{equation}
for $j=1,2$.
The presence of the condensate with $\rho_{j0} \neq0$, which is a temporal and spatial constant, leads to
spontaneous breaking of the $U(1)$ gauge symmetry and results in a
profound consequence for the spectrum of the quasiparticle
excitations, giving the gapless modes of phonons.

The Hamiltonian density of the spacetime-independent condensates is thus obtained from
the Lagrangian density~(\ref{eq:1}) in terms of $\rho_{j0}=N_{j0}/V$ and $\theta_{j0}$ as
\begin{align}
\mathcal{H}_0=C\rho_0 +\frac{1}{2}\Delta\rho_0z-\frac{1}{4}A\rho_0z^2+\frac{1}{2}\Omega\rho_0\sqrt{1-z^2}\cos{\theta_{12}}, \label{Ham}
\end{align}
where the new variables, the density difference and phase difference, are introduced as
\be
z\equiv (\ra-\rb)/(\ra+\rb)  \, ,\quad  \theta_{12}=\theta_{10}-\theta_{20} \,
\ee
with total density $\rho_0=\rho_{10}+ \rho_{20}$.
The density difference $z$ is normalized in a way that for $z=1 (-1)$, all atoms are in condensate 1 (condensate 2).
In~(\ref{Ham}), we have introduced the parameters of the mean-field energy as
\begin{align}
&C\equiv (2g_{12}+g_{11}+g_{22})\rho_0/8,\\[6pt]
&\Delta \equiv \left[(g_{11}-g_{22}  )\rho_0- 2\delta\,\right]/2,\\[6pt]
&A\equiv (2g_{12}-g_{11}-g_{22})\rho_0/2.
\end{align}
%
%Also, $C\equiv (2g_{12}+g_{11}+g_{22})\rho_0/8$, $\displaystyle \Delta \equiv \left[(g_{11}-g_{22}  )\rho_0- %2\delta\,\right]/2$, and $ A\equiv (2g_{12}-g_{11}-g_{22})\rho_0/2$.
%
The quantity of $C$ contributes to the shift of the energy per atom.
 Moreover, we can treat $\Delta$ as an effective detuning parameter, given in terms of the detuning parameter $\delta$ and the difference between the coupling constants within each hyperfine state of the atoms.
Notice that the nonvanishing value of  $\Delta$  explicitly breaks the symmetry of $z \rightarrow -z$ in the system, resulting in
the lowest energy state with $z\neq0$ ( $\rho_{10} \neq \rho_{20}$ ).
Nevertheless, $A$ can be regarded as an effective mutual interaction between two different hyperfine states. For example, $A<0$ implies that the interaction strength of the atoms between different states is smaller than  the average of interaction strengths  within the same state, $g_{12} < (g_{11}+g_{22})/2$, so that the overlap of  the condensates 1 and 2 is energetically favored, resulting in relatively lower population imbalance. In the situation of $A >0$, the system instead favors a high population imbalance ($\vert z \vert \rightarrow \pm1$).
Minimizing the Hamiltonian density~(\ref{Ham}) with respect to $z$ and $\theta_{12}$  with a fixed total number density
$\rho_0=\rho_{10}+\rho_{20}$ gives the mean-field equations
\begin{eqnarray}
&& \sin\theta_{12}=0 \, ,\\
&& -\Delta + Az +  \Omega\frac{z}{\sqrt{1-z^2}}\cos{\theta_{12}} =0 \, .\label{zeqm}
\end{eqnarray}
The phase difference $\theta_{12}=\pi$ is obtained for  the lowest energy state of the Hamiltonian density in~(\ref{Ham}).
The mean-field equation is solved numerically with the given set of the parameters $( g_{11}, g_{12}, g_{22}
, \delta, \Omega)$.

\begin{figure}[t]
\begin{center}
\includegraphics[width=1\linewidth]{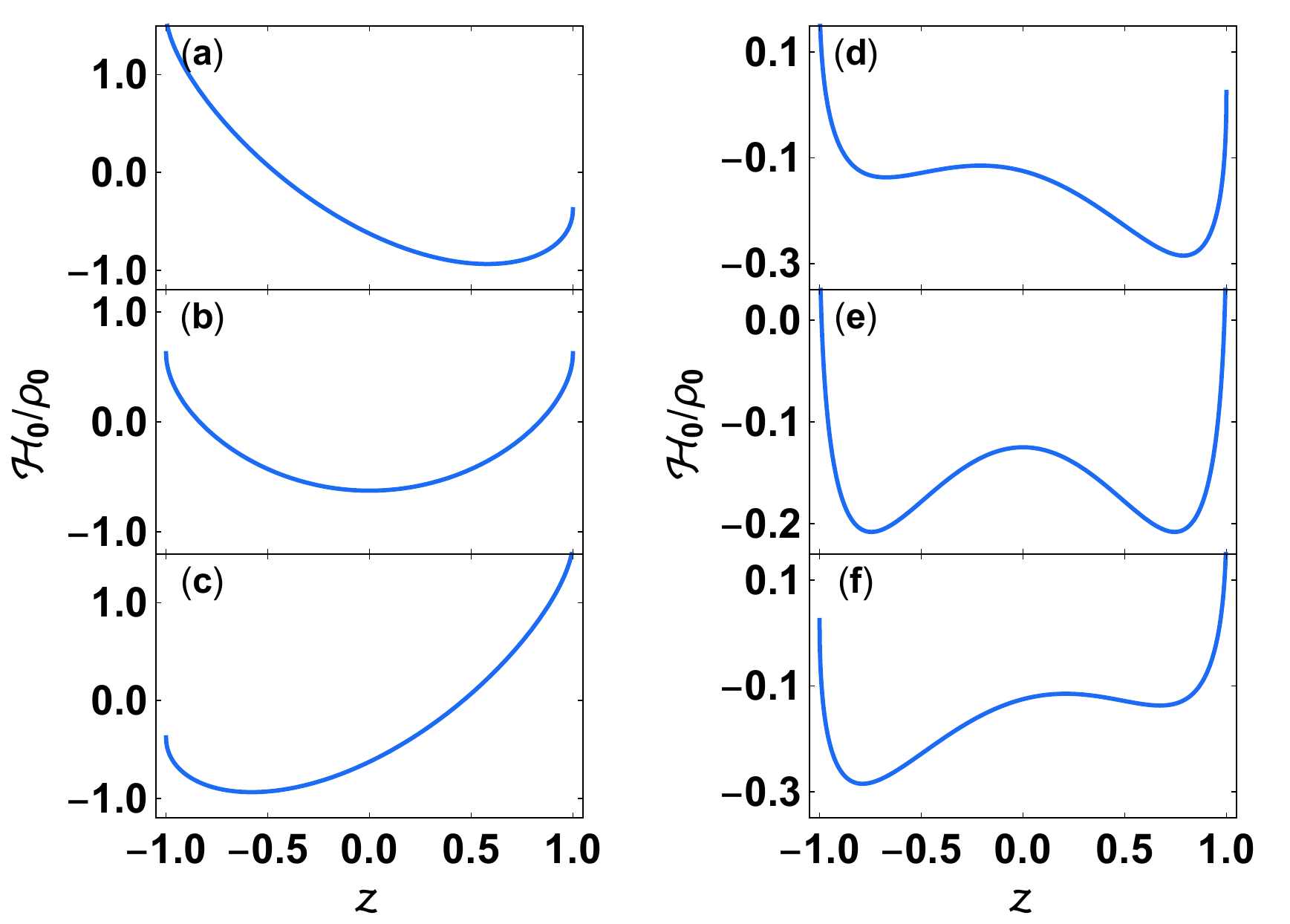}
\caption{ The energy density of the condensate normalized by the total density ($\mathcal{H}_0/\rho_0$) versus the variable $z$. The phase difference $\theta_{12}=\pi$ is obtained for the ground state. The  exemplary values are $g_{11} \rho_{0} =g_{22}\rho_{0}=1.0 \Omega$ and $g_{12}\rho_{0}=0.5 \Omega$ so that $A/\Omega= -0.5 <1$.  The detuning parameter $\delta$ is also chosen to give
$\Delta=-1.0 \Omega$ in (a), $\Delta=0$ in (b), and  $\Delta=1.0 \Omega$ in (c).  The other choices are $g_{11} \rho_0 =g_{22} \rho_0 =1.0 \Omega$ and $g_{12} \rho_0 =2.5\Omega$ so that $A/\Omega=1.5 >1$. Tuning the $\delta$ gives
 $\Delta=-0.1\Omega$ in (d), $\Delta=0$ in (e) , and
 $\Delta = 0.1\Omega$ in (f), with $\vert \Delta \vert < \Delta_c=(A^{2/3}-\Omega^{2/3})^{3/2}$ where the system has one metastable  state and one unstable state, and as well as the ground state. }
\label{mean_field_H_A}
\end{center}
\end{figure}

\begin{figure}[t]
\begin{center}
\includegraphics[width=1\linewidth, trim={0 0 0.7cm 1cm},clip]{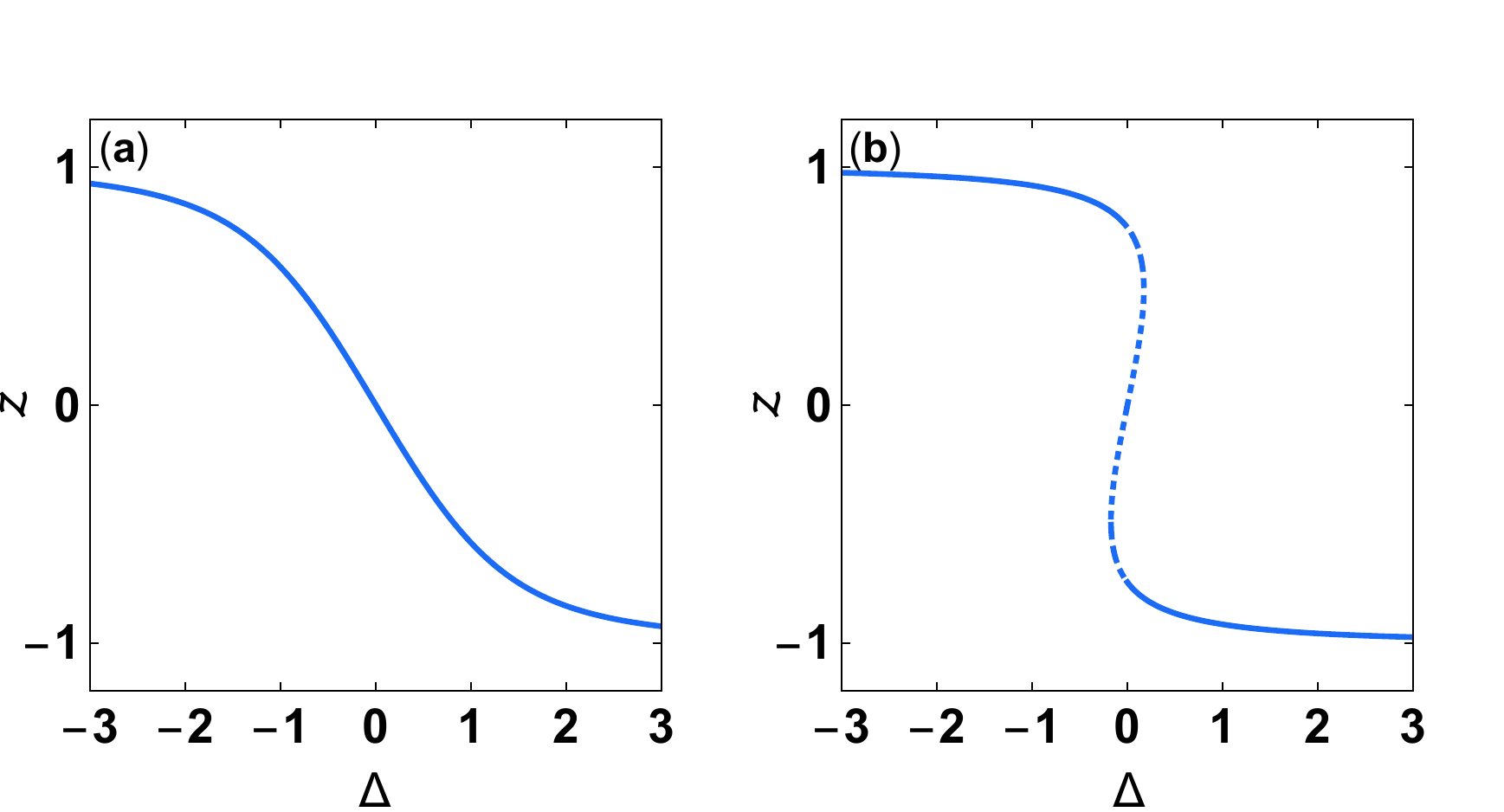}
\caption{(a) The population imbalance $z$,
resulting from solving the mean-field equation, is found to have, for $A/\Omega <1$, the continuous change from $z \rightarrow 1$ ($\rho_{10} \gg \rho_{20} $) to $ z\rightarrow -1$ ($\rho_{20} \gg \rho_{10}$) by changing $\Delta$ with the values from Figs.\,\ref{mean_field_H_A}(a) to \ref{mean_field_H_A}(c). (b) For $A/\Omega >1$, $z$ exhibits the discontinuous change when  $\Delta$ is tuned with the values from Figs.\,\ref{mean_field_H_A}(d) to \ref{mean_field_H_A}(f). In addition, the dotted lines denote  population imbalance $z$ in the  metastable/unstable states as a function of $\Delta$. }
\label{mean_field_z_A}
\end{center}
\end{figure}

In general, when $A/\Omega <1$, there exists the global energy minimum state in the system.
 In Figs.\,\ref{mean_field_H_A}(a)--\ref{mean_field_H_A}(c), we show $\mathcal{H}_0/\rho_0$ as a function of $z$ by choosing the parameters for $A/\Omega <1$ and $\Delta <0$ in (a), $\Delta =0$ in (b), and $\Delta >0$ in  (c), respectively.
 Fig.\,\ref{mean_field_z_A}(a) illustrates the ground state solution of the system for $A/\Omega <1$. Notice that, as $\Delta$ changes from $\Delta <0$ to $\Delta >0$, $z$  changes continuously from the $z>0$ phase   to the $z<0$ phase.
 Also, for $A/\Omega <1$, the mean-field equations can be solved analytically as
\begin{align} \label{z_a}
z &\simeq \frac{\Delta}{A-\Omega}+{\cal{O}} (\Delta^2)\, ,  \hspace{2.3cm}\text{for}\quad\Delta \rightarrow 0 \,,\no
 z & \simeq 1 - \frac{\Omega^2}{2 (\vert\Delta\vert+A)^2} \,+{\cal{O}} (1/\Delta^3) , \hspace{0.5cm}\text{for}\quad\Delta \ll 0 \,, \no
z &\simeq  -1+ \frac{\Omega^2}{2( \vert\Delta\vert+A)^2} \,-{\cal{O}} (1/\Delta^3) , \hspace{0.2cm}\text{for}\quad \Delta \gg 0 \, ,
\end{align}
consistent with Fig.\,\ref{mean_field_z_A}(a).
 By choosing $A/\Omega >1$ with the mean-field results shown in Figs.\,\ref{mean_field_H_A}(d)--\ref{mean_field_H_A}(f), in addition to the ground  state, there exists one metastable state  and one unstable state in the system,  as long as $ \Delta  < \vert \Delta_c \vert$, for $\Delta <0$ in \ref{mean_field_H_A}(d) or $\Delta >0$ in \ref{mean_field_H_A}(f). The ground states become degenerate for  $\Delta =0$ in \ref{mean_field_H_A}(e).
When  $\Delta=\Delta_c= \left(A^{2/3}-\Omega^{2/3}\right)^{3/2}$,
the metastable state solution merges with the unstable state solution. Thus, for $A/\Omega > 1$ but $ \Delta  > \vert \Delta_c \vert$, the system then has the global energy minimum state only.
In Fig.\,\ref{mean_field_z_A}(b), which summarizes the ground state solutions for $A/\Omega > 1$,  the population imbalance $z$ of the condensates changes discontinuously from the condensate 1 dominated to the condensate 2 dominated by changing $\Delta$ from $\Delta >0$ to $\Delta < 0$ and vice versa.
 The analytical mean-field solutions  in the  case of $A/\Omega >1$  are found to be
\begin{align}\label{z_A}
z & \simeq \sqrt{1- \Omega^2/A^2} -{\cal{O}} (\Delta) \, ,   \hspace{1.38cm}\text{for}\quad\Delta \rightarrow 0^{-}\, , \no
z & \simeq -\sqrt{1- \Omega^2/A^2} +{\cal{O}} (\Delta) \,   , \hspace{1.1cm}\text{for}\quad\Delta \rightarrow 0^{+} \, ,
\end{align}
and for $\Delta\gg 0$ and $\Delta \ll 0$, the solutions have the same approximate forms in (\ref{z_a}).
\begin{figure}[t]
\begin{center}
\includegraphics[width=1\linewidth]{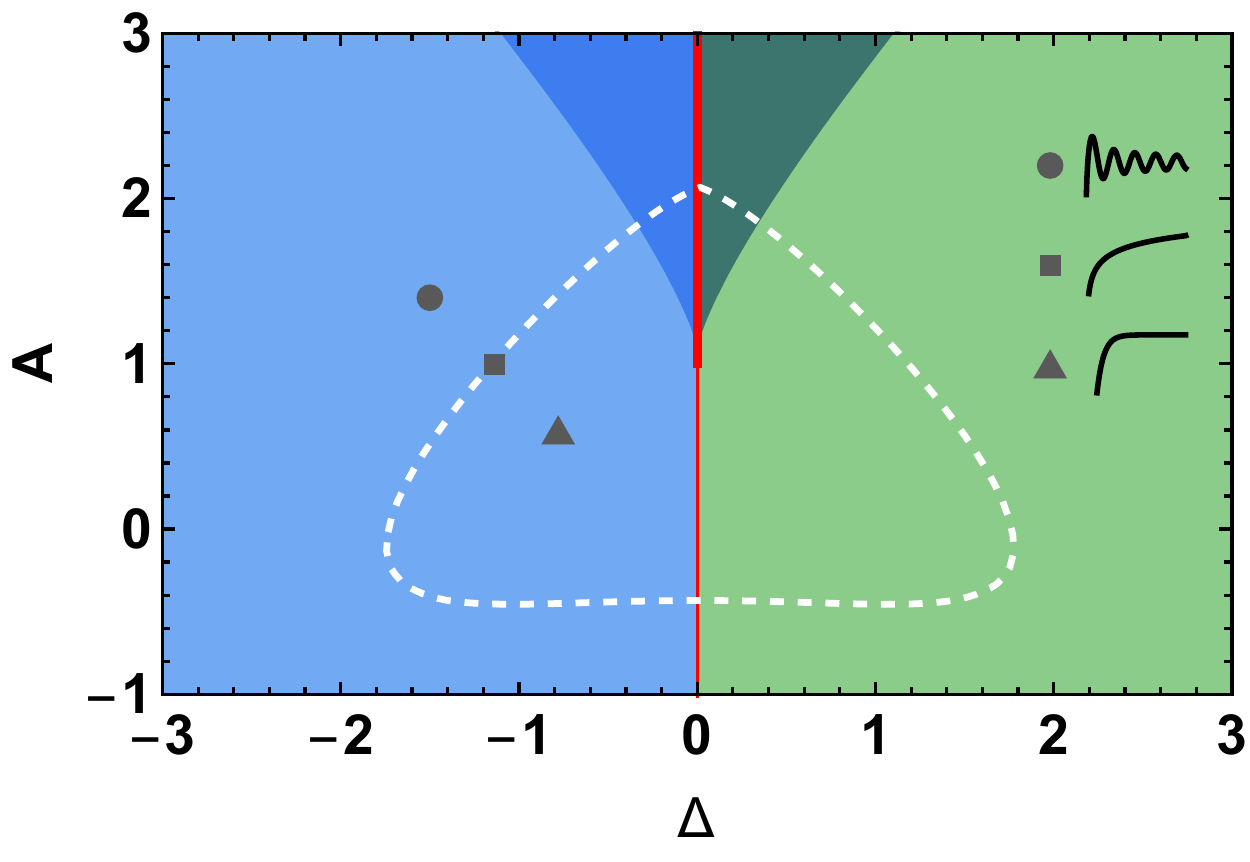}
\caption{The diagram  shows the regimes of the parameters ($A$, $\Delta$) in distinct phases.
 In the conditions of $m_1=m_2=m$ and $g_{11}=g_{22}$ as an example, the parameters in the blue regime are for the  $z>0$ phase  whereas in the green regime they are for the  $z<0$ phase.
 Moreover, the parameters in the dark-colored regime  (for both the blue and green regimes) lead to the system with one metastable state and one unstable state in addition to the global ground state, whereas the parameters in the light-colored
 regime are for the system with the ground state only.
 Population imbalance $z$ changes discontinuously from one phase to the other as crossing the boldfaced solid (red) line whereas  $z$ changes continuously  across two phases through the (red) line.  In addition, the white dotted line shows the values of $A$ and $\Delta$, which give  the same speeds of the gapless and the gapped modes.  Three marks with different geometric shapes show the values of $A$ and $\Delta$  by choosing $a_{12}$ and $\delta$ that exhibit three different behaviors of TOF variance respectively as a function of the flight time. }
\label{mean_field_summary}
\end{center}
\end{figure}
The diagram in~Fig.\,\ref{mean_field_summary} shows distinct phases (the phases with $z>0$ and $z<0$) and different types of the mean-field solutions in terms of the parameters $\Delta, A$ by tuning $( g_{11}, g_{12}, g_{22}
, \delta)$.
In short, the population imbalance $z$ of the binary condensates changes discontinuously as $\Delta$ goes from $\Delta <0$ to $\Delta >0$ for $A/\Omega>1$, whereas $z$ changes continuously across two phases for $A/\Omega<1$. Thus, the discontinuous change in $z$ across two phases is terminated when $A/\Omega =1$.

The idea of using the one-component BEC system for building up the analogue gravity model has been applied intensively to mimic various phenomena in general relativity and in the early Universe geometries.
Nevertheless, it is seen that the two-component model  have richer phase structures,
which can be adopted to stimulate such as the dynamics of the (continuous/discontinuous)  phase transitions
of the early Universe on the emergent spacetime, mimic by the time- and/or  space-dependent Bose condensates.
\vspace{1cm}
\section{the dispersion relation for collective excitations} \label{sec2}
We proceed by perturbing the system around the condensates and try to explore the dynamics of the collective excitations.
To do so, the field is expanded around its mean-field value by
\begin{align}
\psi_j (\mathbf{x},t)=\sqrt{\rho_{j0}+\delta\rho_j (\mathbf{x},t)}e^{i(\theta_{j0}+\delta\theta_j (\mathbf{x},t))} \, .
\end{align}
The Lagrangian density  can split into
\begin{align}
\mathcal{L}=\mathcal{L}_0 [\rho_{j0}, \theta_{j0}]+ \mathcal{L}_{\rm fluct} [\delta \rho_{j},\delta\theta_{j}]\, ,
\end{align}
 where $\mathcal{L}_0$ is the Lagrangian density of the background condensates with the corresponding mean-field Hamiltonian density  $\mathcal{H}_0$ given in (\ref{Ham}), and
$ \mathcal{L}_{\rm fluct} [\delta\rho_{j},\delta\theta_{j}]$ is the Lagrangian density, governing the dynamics of the fluctuations of
the condensates.
\begin{comment}
\begin{align}
\mathcal{L}_0 = \sum_{i=1}^2\bigg{\{}-\rho_{i0}\Big{[}\,\dot{\theta}_
i+\frac{(\nabla\theta_{i0})^2}{2m_i}\Big{]}-\frac{(\nabla\rho_{i0})^2}{8m_i\rho_{i0}}+(\mu_i-\frac{\delta_i}{2})\rho_{i0}-\frac{g_{ii}\rho_{i0}^2}{2}\bigg{\}}
-g_{12}\,\ra\rb+\Omega\,\sqrt{\ra\rb},
\end{align}
\end{comment}
We expand $ \mathcal{L}_{\rm fluct} [\delta\rho_{j},\delta\theta_{j}]$  in the { small} $\delta\theta_j$ and the fluctuations in the condensate
density
 $\delta\rho_j$, expressing all terms  obeying the Galilean invariance of the system. Immediate Galilean invariants of the theory are the
density fluctuation
 $\delta \rho_j$ itself, and
$G(\delta\theta_j) = {\delta\dot{\theta}_j} + (\nabla \delta\theta_j )^2/2 m_j$.
The resulting Galilean invariant effective action
 up to the cubic terms in the field variables takes the form
\begin{align}
\label{lag}
&\mathcal{L}_{\text{fluct}}[\delta\rho_{j}, \delta\theta_{j}]=-G(\delta\theta_1)\,\delta\rho_1-G(\delta\theta_2)\,\delta\rho_2\no
&-\frac{(\nabla\delta\rho_1)^2}{8m_1\ra}-\frac{(\nabla\delta\rho_2)^2}{8m_2\rb}-\lambda_{11}\delta\rho_1^2-\lambda_{22}\delta\rho_2^2-\lambda_{12}\delta\rho_1\delta\rho_2\no
&-\frac{\ra}{2m_1}G(\delta\theta_1)^2-\frac{\rb}{2m_2}G(\delta\theta_2)^2-\frac{\Omega}{2}\sqrt{\ra\rb}\,(\delta\theta_1-\delta\theta_2)^2,
\end{align}
where we have introduced the effective interactions
\begin{align}
&\lambda_{11}\equiv\frac{g_{11}}{2}+\frac{\Omega}{8}\sqrt{\frac{\rb}{\ra}}\frac{1}{\ra},\no
&\lambda_{22}\equiv\frac{g_{22}}{2}+\frac{\Omega}{8}\sqrt{\frac{\ra}{\rb}}\frac{1}{\rb},\no
&\lambda_{12}\equiv g_{12}-\frac{\Omega}{4}\frac{1}{\sqrt{\ra\rb}}. \label{lambdadef}
\end{align}
Apart from the quadratic terms, to be diagonalized to obtain the dispersion relation of the quasiparticles, the Galilean invariant, $G(\delta\theta_j)$, generates the coupling with the cubic terms which will be treated later perturbatively.

The next step is to notice that the field
$\delta \rho_j$ does not have its time derivative term in $\mathcal{L}_{\text{fluct}}$~(\ref{lag}), which can be readily eliminated using its equation of  motion
%obtained by taking the variation of $ \mathcal{L}_{\text{fluct}}$ with respect to $\delta\rho_{1,2}$ given by
%
\begin{align}
&-\frac{1}{4m_1\ra}(\nabla^2\delta\rho_1)+2\lambda_{11}\delta\rho_1+\lambda_{12}\delta\rho_2+G(\delta\theta_1)=0\, ,\no
&-\frac{1}{4m_2\rb}(\nabla^2\delta\rho_2)+2\lambda_{22}\delta\rho_2+\lambda_{12}\delta\rho_1+G(\delta\theta_2)=0.
\end{align}
Substituting the solutions back into $\mathcal{L}_{\text{fluct}}$~(\ref{lag}),
%the Lagrangian to replace the densities $\delta\rho_j$,
and retaining the terms up to fourth order derivatives, we have the Lagrangian density in the quadratic form
\begin{align}
\label{L_q}
&\mathcal{L}_{\text{q}}=\no&-\frac{1}{8\bl_0^2}\left(\frac{4\lb^2}{m_1\ra}+\frac{\lab^2}{m_2\rb}\right)(\nabla\delta\dot{\theta}_1)^2-\frac{\ra}{2m_1}(\nabla\delta\theta_1)^2\no&-\frac{1}{8\bl_0^2}\left(\frac{4\la^2}{m_2\rb}+\frac{\lab^2}{m_1\ra}\right)(\nabla{\delta\dot{\theta}_2})^2-\frac{\rb}{2m_2}(\nabla\delta\theta_2)^2\nonumber\\
&+\frac{\lab}{2\bl_0^2}\left(\frac{\la}{m_2\rb}+\frac{\lb}{m_1\ra}\right)(\nabla \delta\dot\theta_1)\cdot(\nabla\delta\dot\theta_2)\nonumber\\
&+\frac{\lb}{\bl_0}{\delta{\dot{\theta}_1}}^2+\frac{\la}{\bl_0}{\delta{\dot{\theta}_2}}^2-\frac{\lab}{\bl_0}\delta\dot{\theta}_1\delta\dot{\theta}_2\no
&-\frac{\Omega}{2}\sqrt{\ra\rb}\,(\delta\theta_1-\delta\theta_2)^2,
\end{align}
where
$\bar\lambda_0=4 \lambda_{11} \lambda_{22} -\lambda_{12}^2$. The cubic terms read
\begin{align}
\label{L_c}
\mathcal{L}_\text{c}=&\,\,\,\,\,
\frac{\lb}{\bl_0}\frac{\delta\dot{\theta}_1(\nabla\delta\theta_1)^2}{m_1}+\frac{\la}{\bl_0}\frac{\delta\dot{\theta}_2(\nabla\delta\theta_2)^2}{m_2}\no
&-\frac{\lab}{\bl_0}\bigg{[}\frac{\delta\dot{\theta}_1(\nabla\delta\theta_2)^2}{2m_2}+\frac{\delta\dot{\theta}_2(\nabla\delta\theta_1)^2}{2m_1}\bigg{]},
\end{align}
as a consequence of the coupling term $G(\delta\theta_1) G(\delta\theta_2)$.

To be more compact, let us express $\mathcal{L}_{\text{q}}$ in a matrix notation
\begin{align}
\mathcal{L}_\text{q}=\Theta \,\Pi\,\Theta^\textbf{T},
\end{align}
 with the vector $\Theta=(\delta\theta_1,\,\,\delta\theta_2)$. The Fourier transformed field $\tilde \theta_j$ is defined as
 \be
 \delta\theta_j(x)= \int \frac{d^{D+1} k}{ (2 \pi)^{D+1}} \, \tilde \theta_j (k) \, e^{-ik x} \, .
 \ee
The  elements of the $\Pi$ matrix can be read off from~(\ref{L_q}) in  momentum space as
\begin{align}
\Pi_{11}=&-\frac{1}{8\bl_0^2}\left(\frac{4\lb^2}{m_1\ra}+\frac{\lab^2}{m_2\rb}\right)k^2\omega^2+\dfrac{\lambda_{22}}{{\bl_0}}\omega^2-\dfrac{\ra}{2m_1}k^2\no
&-\dfrac{\Omega\sqrt{\ra\rb}}{2},\no
\Pi_{12}=&\frac{\lab}{4\bl_0^2}\left(\frac{\la}{m_2\rb}+\frac{\lb}{m_1\ra}\right)k^2\omega^2-\dfrac{\lambda_{12}}{2\bl_0}\omega^2+\dfrac{\Omega\sqrt{\ra\rb}}{2},\no
\Pi_{21}=&\frac{\lab}{4\bl_0^2}\left(\frac{\la}{m_2\rb}+\frac{\lb}{m_1\ra}\right)k^2\omega^2-\dfrac{\lambda_{12}}{2\bl_0}\omega^2+\dfrac{\Omega\sqrt{\ra\rb}}{2},\no
\Pi_{22}=&-\frac{1}{8\bl_0^2}\left(\frac{4\la^2}{m_2\rb}+\frac{\lab^2}{m_1\ra}\right)k^2\omega^2+\dfrac{\lambda_{11}}{{\bl_0}}\omega^2-\dfrac{\rb}{2m_2}k^2\no
&-\dfrac{\Omega\sqrt{\ra\rb}}{2}.
\end{align}
We decouple the modes $\tilde\theta_j$ for $j=1,2$ by diagonalizing $\mathcal{L}_{\text{q}}$
%the  quadratic Lagrangian, and
using the Bogoliubov transformation \cite{tim,lin06},
% the eigenvectors can be written as
\begin{align}
\chi_{\omega,k} &=\cos{\gamma_{\omega,k}}\,\tilde \theta_1-\sin{\gamma_{\omega,k}}\,\tilde \theta_2, \no
\phi_{\omega,k} &=\sin{\gamma_{\omega,k}}\,\tilde \theta_1+\cos{\gamma_{\omega,k}}\,\tilde \theta_2,
\end{align}
where
\begin{align}
\cos{2\gamma_{\omega,k}}=\dfrac{\Pi_{11}-\Pi_{22}}{\left[(\Pi_{11}-\Pi_{22})^2+4{\Pi_{12}}^2\right]^{1/2}}.
\end{align}
%\vspace{0.5cm}
%gives
% $
%\sin{\gamma_{\omega,k}} = \big{(}1-\cos{2\gamma_{\omega,k}}\big{)}^{1/2}/\sqrt{2}$ and $\cos{\gamma_{\omega,k}} = %\big{(}1+\cos{2\gamma_{\omega,k}}\big{)}^{1/2}/\sqrt{2}$.
In terms of the eigenvectors $\phi$ and $\chi$, the Lagrangian density $\mathcal{L}_{\text{q}}$ becomes
%then can be rewritten as two decoupled modes, given by
\begin{align}
\mathcal{L}_\text{q} = \bar\Omega_\phi\,\phi^\dagger\phi+\bar\Omega_\chi\,\chi^\dagger\chi \, .
\end{align}

The eigenvalues $\bar\Omega_\phi$ and $\bar\Omega_\chi$
%$\bar{\Omega}_{\pm}$
are determined from the secular equation
%following determinant:
\begin{align}
\begin{array}{|cc|}
    \Pi_{11}-\bar{\Omega} & \Pi_{12} \\
    \Pi_{21} & \Pi_{22}-\bar{\Omega}
\end{array}=0\;.
\end{align}
The calculations for the quasiparticle energies are lengthy but straightforward. We have found
\begin{widetext}
\begin{align}
\bar\Omega_{\pm}=&\frac{1}{2}\Bigg\{\frac{(\la+\lb)}{\bl_0}\omega^2-(\ra-\gamma\rb)\frac{k^2}{2m_1}-\Omega\sqrt{\ra\rb}\pm\bigg[(\ra-\gamma\rb)^2 \frac{k^4}{4m_1^2}+\frac{2}{\bl_0}\Big((\la-\lab)\ra\no&
\qquad+\gamma(\lb-\lab)\rb\Big)\omega^2 \frac{k^2}{2 m_1}+\frac{1}{\bl_0^2}\Big(\lab^2+(\la-\lb)^2\Big)\omega^4-\frac{2\lab}{\bl_0}\sqrt{\ra\rb}\,\Omega\,\omega^2+\ra\rb\Omega^2\bigg]^{1/2}
\Bigg\},
\end{align}
\vspace{0.5cm}
\end{widetext}
where $\gamma=m_1/m_2$ is the mass ratio, and the gapless and gapped energy spectra are given, respectively, by $\bar\Omega_\phi=\bar\Omega_-$ and  $\bar\Omega_\chi=\bar\Omega_+$.
In the limit of small $\omega$ and $k$, we find
\begin{align} \label{Omega_phi}
\bar{\Omega}_\phi =\,\frac{\Lambda_{\phi}}{2} \,\omega^2-\frac{\rho_{\phi}}{4m_1}k^2 -\frac{ 2 \zeta_{\phi}}{ \Lambda_{\phi}} k^4,
\end{align}
for the gapless mode (the phonon).
As for $\bar\Omega_{\chi}$,  it is expanded around the gap energy to be determined from its on-shell condition.
We write the leading terms as
% and in the small $k$ approximation given by
\begin{align} \label{Omega_chi}
\bar{\Omega}_\chi =\,\frac{\Lambda_{\chi}}{2} \,\omega^2-\frac{\rho_{\chi}}{4m_1}k^2 -\frac{M^2}{2}\, .
\end{align}
All the coefficients in (\ref{Omega_phi}) and (\ref{Omega_chi}) are listed below:
\begin{align}
&\Lambda_{\phi} =\frac{\la+\lb-\lab}{4\la\lb-\lab^2} , \no[6pt]
&\Lambda_{\chi} = \frac{\la+\lb-\lab}{ 2\la^2-2\la\lab+\lab^2-2\lab\lb+2\lb^2}\, ,\no[6pt]
&\rho_{\phi}=\ra+\gamma\rb\, ,\no[6pt]
&\rho_{\chi} = \frac{ (\lab-2\la)^2\ra+(\lab-2\lb)^2 \,\gamma \rho_{20}}{2\la^2-2\la\lab+\lab^2-2\lab\lb+2\lb^2 }
 \, , \nonumber\\[6pt]
&M^2 =\frac{2 \Omega\sqrt{\ra\rb} \, (\la+\lb-\lab)^2 }{2\la^2-2\la\lab+\lab^2-2\lab\lb+2\lb^2} \, , \label{coeffs}
\end{align}
\begin{align}
\zeta_\phi\,
 =&-\frac{1}{\Omega}\frac{\gamma^2}{8\,m_1^2}\frac{1}{\sqrt{\ra\rb}}\frac{4\la\lb-\lab^2}{(\la+\lb-\lab)^3}\no
 &\times[\rb(\lab-2\lb)-\gamma\ra(\lab-2\la)]^2\no
&+\frac{\gamma^2}{16m_1^2}\frac{1}{\ra\rb}\frac{\rb+\gamma\ra}{(\la+\lb-\lab)^2}\no
&\times[\ra(\lab-2\la)^2+\gamma\rb(\lab-2\lb)^2] \, .  \label{zetaphi}
%\zeta_\chi \,=\,&\frac{1}{\Omega}\frac{\gamma^2}{8\,m_1^2}\frac{1}{\sqrt{\ra\rb}}\frac{4\la\lb-\lab^2}{(\la+\lb-\lab)^3}\no
%&\times[\rb(\lab-2\lb)-\gamma\ra(\lab-2\la)]^2\no
%& +\frac{\gamma^2}{16m_1^2}\frac{1}{\ra\rb}\frac{1}{(\la+\lb-\lab)^2}\no
%&\times\Big\{\ra\rb(\lab-2\la)(-4\la+3\lab-2\lb)\no
%&-\gamma[\ra^2(\lab-2\la)^2+\rb^2(\lab-2\lb)^2]+\no
%&\gamma^2\ra\rb(\lab-2\la)(-2\la+3\lab-4\lb)\Big\}.\label{coeffs}
\end{align}
The dispersion relations of the modes $\phi$ and $\chi$ are obtained at their on-shell conditions by taking $\bar \Omega_{\phi, \chi} =0$ in (\ref{Omega_phi}) and (\ref{Omega_chi}). With the help of (\ref{coeffs}) and (\ref{zetaphi}) we have
%and, in the small $k$ approximation, expressed as
\begin{align}\label{cs}
&\omega_\phi^2=c_{\phi}^2 k^2 + \zeta_{\phi}k^4\, ,\\[6pt]
&\omega_\chi^2=m_\text{eff}^2+c_{\chi}^2k^2 \,.
\end{align}
% In addition to the gapless $\phi$ mode,
%there is the gapped modes of the $\chi$ field  with the effective mass
The gapped mode $\chi$ has an effective mass given by
\bea
m_\text{eff}^2=\frac{M^2}{\Lambda_{\chi}}= 2 \Omega\sqrt{\ra\rb} (\la+\lb-\lab) \, .
\eea
The speeds of the modes are
\bea \label{c_2}
c_{\phi}^2 &=& \frac{\rho_{\phi}} { 2 m_1 \Lambda_{\phi}}  = \frac{\ra+\gamma\rb}{2m_1}\bigg(\frac{4\la\lb-\lab^2}{\la+\lb-\lab}\bigg),\\[6pt]
c_{\chi}^2 &=& \frac{\rho_{\chi}} { 2 m_1 \Lambda_{\chi}} =\frac{(\lab-2\la)^2\ra+(\lab-2\lb)^2\gamma\rb}{2 m_1(\la+\lb-\lab)} \, .\no\label{c_22}
\eea
It is worthwhile to mention that the behaviors of the speeds, the effective mass and the coefficient of the $k^4$ term are all given in terms of the parameters of the model and the mean-field solutions discussed earlier in Sec II. In the following, we shall study their limits in the regime of $\Delta \rightarrow 0$ and $\Delta \gg 0$, $\Delta \ll0$.
%
%The analytical properties of above quantities
Detailed control of these coefficients find themselves to be very helpful for digging out the appropriate values of the tunable parameters in order to get large sound cone fluctuations, which is the main goal of the present work.
%These two dispersion relations are in the desired form that only permits even powers in $k$.
%The behaviors of the speeds of two modes, the effective mass as well as the coefficient of the $k^4$ term can be %analyzed analytically in the regimes of $\Delta \rightarrow 0$ and $\Delta \gg 0$, $\Delta \ll0$.

For $A/\Omega<1$, we find that the speeds $c_{\phi}$ and $c_{\chi}$ go with $\Delta$ as
\vspace{0.2cm}
\begin{align}\label{cphi_a}
c^2_{\phi}\simeq &\,\,{(1+\gamma)[2(g_{11} g_{22}-g_{12}^2)\rho_0^2 +(g_{11}+g_{22}+2 g_{12})\Omega \rho_0]}\no
&\,/{8 m_1 (\Omega-A)} +{\cal{O}}(\vert \Delta \vert), \hspace{1.45cm} {\text{for}}\quad  \Delta \rightarrow 0,\no[7pt]
c_{\phi}^2 \simeq &\,\, \frac{g_{22} \gamma \rho_0}{m_1} +{\cal{O}}(1/\Delta^2),\hspace{2cm} {\text{for}} \quad \Delta \gg 0,\no
c_{\phi}^2 \simeq  &\,\, \frac{g_{11}  \rho_0}{m_1} +{\cal{O}}(1/\Delta^2),\hspace{2.2cm}\text{for} \quad\Delta \ll 0\,,
\end{align}
\begin{align}
c_{\chi}^2\simeq &\,\,\{(1+\gamma) \Omega^2 +[(g_{11}-g_{12})^2+\gamma (g_{12}-g_{22})^2]\rho_0^2\hspace{1.cm}\no
&+[ g_{11}-(\gamma+1) g_{12} -\gamma g_{22}] \Omega \rho_0\}/8 m_1 (\Omega-A)\no
&+{\cal{O}} (\vert \Delta \vert),\no
&\hspace{4.8cm}\text{for} \quad \Delta \rightarrow 0, \nonumber\\
c_{\chi}^2 \simeq &\,\,\frac{(\vert \Delta \vert +A)}{ 2 m_1} +{\cal{O}} (1/\Delta^2), \hspace{1.2cm}\text{for} \quad  \Delta \gg 0 , \Delta \ll 0.\label{cchi_a}
\end{align}
The speeds behave differently with the detuning parameter $\Delta$. Whereas the sound speed $c_{\phi}$ reaches a finite value in the limit of large $\vert \Delta \vert$, the speed of the gapped mode $c_{\chi}$
grows linearly with $\vert \Delta \vert$.
%It will be shown later
We anticipate here that the tunability of the speeds of the quasiparticles plays an important role
in determining the TOF variance of sound waves influenced by the quantum fluctuations of the gapped modes.

The effective mass $m_{\text{eff}}$ of the gapped mode in this case of $A/\Omega <1$ takes the approximate forms
\begin{align}
m^2_{\text{eff}} \simeq \,\,& \Omega(\Omega-A) + {\cal{O}}(\Delta),\hspace{0.15cm}\quad {\text{for}} \quad  \Delta \rightarrow 0, \label{meff_a_Delta0}\\
 m^2_{\text{eff}} \simeq \,\,& (\vert\Delta\vert+A)^2 + {\cal{O}} (\Delta^0), \hspace{0.2cm}  {\text{for}} \quad  \Delta \gg 0 ,\, \Delta \ll 0.
\label{meff_a}
\end{align}
In the limit of  $\Delta \rightarrow 0 $, we have the mean-field solution $\rho_{10}=\rho_{20} =\rho_0/2$, and also $m^2_{\text{eff}}  \rightarrow 0$ as $\Omega \rightarrow 0$ as shown in~(\ref{meff_a_Delta0}), the gap energy goes to zero.
Recalling Eq.(\ref{eq:1}) with $\Omega\rightarrow 0$, the Lagrangian density ${\mathcal{L}}(\Omega=0)$ is then invariant under the respective $U(1)$ gauge transformation for each $\psi_j$; both $N_1$ and $N_2$ in~(\ref{n1_n2}) are the conserved quantities. The nonzero $\rho_{10}$ and $\rho_{20}$ break both symmetries, in this case. With the choice of the expectation values of the field operators of $\rho_{10}=\rho_{20} =\rho_0/2$, the system has the second Goldstone mode, thus with two gapless modes in the system~\cite{visser,tim}.
%
%Since the Lagrangian ${\mathcal{L}}(\Omega=0)$ (\ref{eq:1}) by setting $\Omega\rightarrow 0$
%is invariant under the respective $U(1)$ gauge transformation for each $\psi_j$, both $N_1$ and $N_2$ %in~(\ref{n1_n2}) are the conserved quantities. The nonzero $\rho_{10}$ and $\rho_{20}$ break both symmetries, and %in particular, with the choice of the expectation values of the field operators of $\rho_{10}=\rho_{20} =\rho_0/2$, %the system has the second Goldstone mode, thus with two gapless modes in the system~\cite{visser,tim}.
%
On the other hand, Eq.(\ref{meff_a}) shows that the effective mass $m^2_{\text{eff}}$ increases with $\vert \Delta \vert$.
This is because, for larger values of $\vert \Delta \vert$, the background fields described by~(\ref{Ham}) have the global lowest energy state with $ \vert  z \vert \rightarrow  1$, with a high population imbalance. Thus, the excitations of the gapped modes having a mixture of the fluctuations with the $\theta_1$ field  and the $\theta_2$ field will cost much energy, roughly given by  large $m_{\text{eff}}$~\cite{tim}.
%Additionally,
%This is because for larger value of $\vert \Delta \vert$, the Hamiltonian density in~(\ref{Ham}) receives the large effect from the linear term in $z$, resulting in the global lowest energy state with $ \vert  z \vert \rightarrow  1$ of highly population imbalance.  Thus, it is anticipated that exciting  the gapped modes to have a mixture of the fluctuations with the $\theta_1$ field  and the $\theta_2$ field will cost much energy, roughly given by  large $m_{\text{eff}}$~\cite{tim}.

\begin{figure}[t]
\begin{center}
\includegraphics[width=1\linewidth]{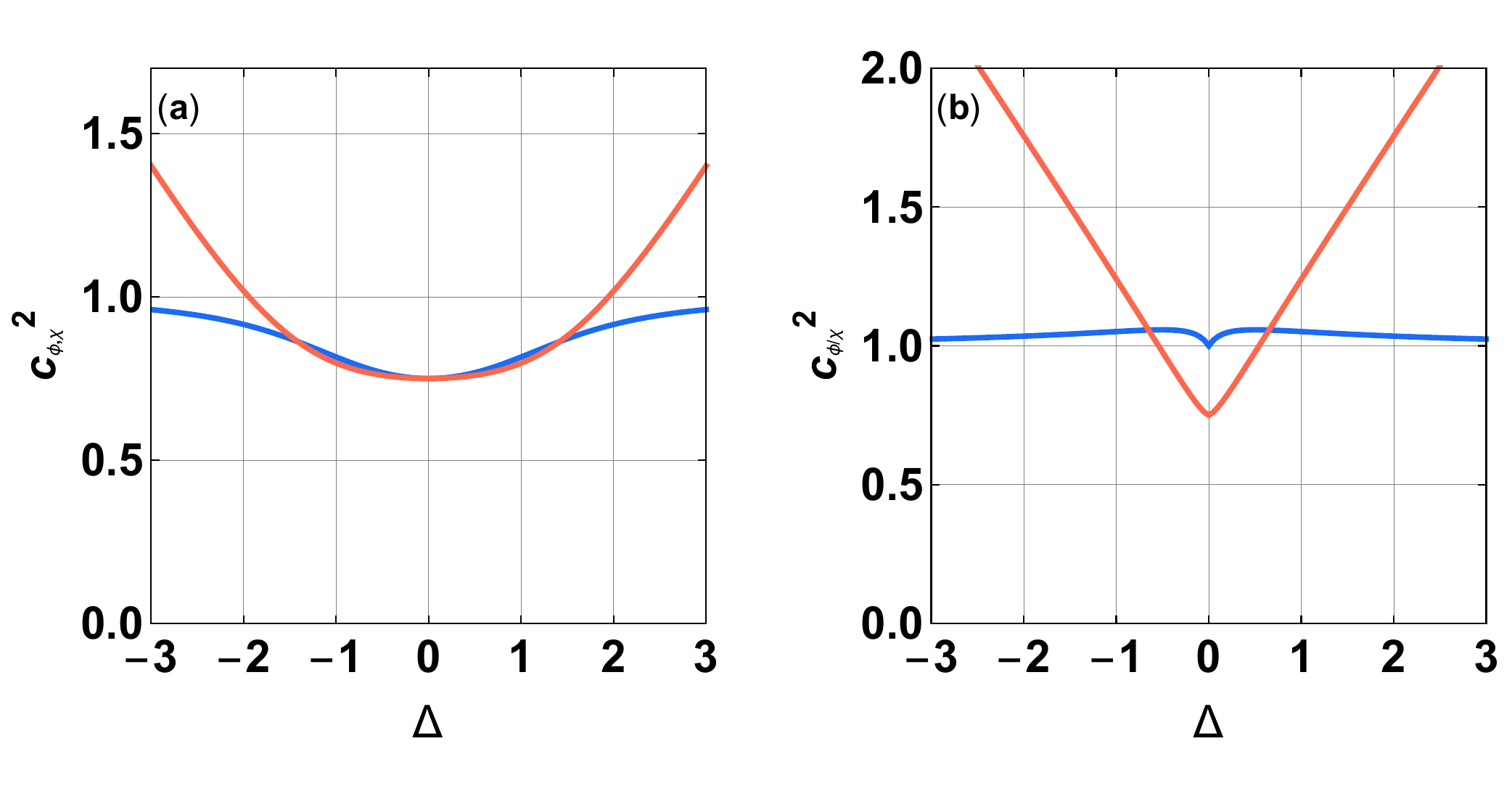}
\caption{The velocity squared in units of $(\Omega/m)^2$ for the  gapless mode (blue) and the gapped mode (red) as functions of  $\Delta$ for the following combinations of parameters:  (a)$g_{11}\rho_0 =g_{22} \rho_0=1.0\Omega$ and $g_{12} \rho_0=0.5\Omega$
for  $A/\Omega <1$ ($A=-0.5\Omega$); (b)$g_{11} \rho_0=g_{22} \rho_0=1.0\Omega$ and $g_{12} \rho_0=2.5\Omega$ for
$A/\Omega >1$ ($A=1.5\Omega$). } \label{c2}
\end{center}
\end{figure}

 For $A/\Omega>1$, since the mean-field solution of $z$ exhibits a discontinuous change from $\Delta \rightarrow 0^+$ to $\Delta \rightarrow 0^-$ in (\ref{z_A}),
the sound speed $c_\phi$ and the velocity for the gapped mode $c_\chi$ will show a similar discontinuity behavior. The equations (\ref{c_2}) and (\ref{c_22}) become, in these limits,
\begin{align}
c_\phi^2 \simeq \,\,& \frac{g_{22}\gamma\rho}{m_1}+\mathcal{O}(|\Delta|),\hspace{1.4cm}\text{for}\quad \Delta\rightarrow 0^+ \, ,\no
c_\phi^2 \simeq\,\, &\frac{g_{11}\rho}{m_1}+\mathcal{O}(|\Delta|),\hspace{1.6cm}\text{for}\quad \Delta\rightarrow 0^-  \, ,
\label{cphi_A}
\end{align}
and
\begin{align}
c_\chi^2 \simeq \,& \frac{A\gamma}{2m_1}+\mathcal{O}(|\Delta|), \hspace{1.3cm}\text{for}\,\, \Delta\rightarrow 0^+ ,\no
c_\chi^2 \simeq \,&  \frac{A}{2m_1}+ \mathcal{O}(|\Delta|),\hspace{1.3cm}\text{for}\,\, \Delta\rightarrow 0^- ,
\label{cchi_A}
\end{align}
whereas for $\Delta \gg 0$ and $\Delta \ll 0$, they show the same approximate results in~(\ref{cphi_a}) and (\ref{cchi_a}), respectively.

As for the effective mass $m^2_{\text{eff}}$ in this case, it is given by
\begin{align} \label{meff_A}
m^2_{\text{eff}} \simeq \,\,& A^2 - \frac{3}{4} \Omega^2+ {\cal{O}}(\Delta),\qquad {\text{for}} \quad  \Delta \rightarrow 0 \, .
\end{align}
\begin{figure}[t]
\begin{center}
\includegraphics[width=1\linewidth]{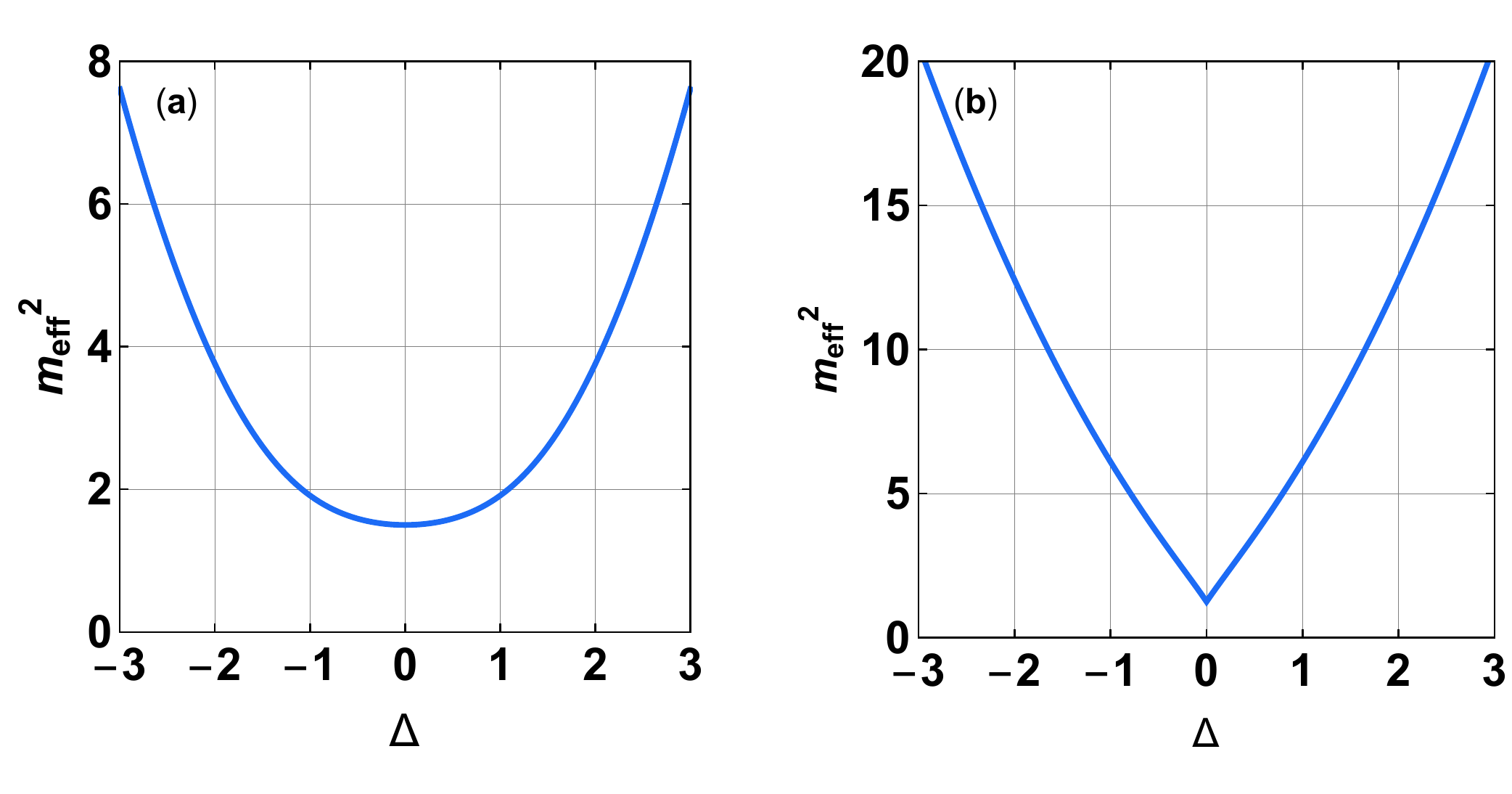}
\caption{The effective mass $m_{\text{eff}}$ as functions of $\Delta$ in the following combinations of parameters:(a)$g_{11}\rho_0=g_{22}\rho_0=1.0\Omega$ and $g_{12}\rho_0=0.5\Omega$ for $A/\Omega <1$ ($A=-0.5\Omega$); (b)$g_{11}\rho_0=g_{22}\rho_0=1.0\Omega$ and $g_{12}\rho_0=2.5\Omega$ for $A/\Omega>1$ ($A=1.5\Omega$).}
\label{meff}
\end{center}
\end{figure}
It is then quite subtle to take the limit $\Omega\rightarrow 0$ in~(\ref{meff_A}) for the interpretation of the second Goldstone mode.
In the case of $A >\Omega=0$, naively setting $\Delta \rightarrow 0$  causes the Hamiltonian density~(\ref{Ham}) with no lower bound. Thus, as long as $\Omega$ is small but remains nonzero, there exist the mean-field solutions $z\rightarrow  \pm 1$ with nonzero $m^2_{\text{eff}}$.
Also for $\Delta \gg 0$ and $\Delta \ll 0$, $m^2_{\text{eff}}$ increases with $\vert \Delta \vert^2$ as in~(\ref{meff_a}).

In general, the speeds of the propagation of two modes are different~\cite{visser}.
While allowing for explicit spatial and/or time dependence in the interaction terms,
the equations of motion for the $\phi$ and $\chi$ fields take the form of the Klein-Gordon equations in curved spacetime, where the acoustic metrics are determined from $c_{\phi}$ and $c_{\chi}$, respectively. These are the local speed of sound, which now  $\Lambda_{\phi}$ and $\Lambda_{\chi}$, etc. vary in space and time. Consequently, the propagations of two modes $\phi$ and $\chi$ experience different metrics.
The condition for having the speeds of two modes equal is when $m_1=m_2=m$, {
 $g_{11}=g_{22}$}, { $\delta=0$ ($\Delta=0$), and $g_{12}=\Omega/4\sqrt{\ra\rb}$ ($\lambda_{12}=0)$ for $A/\Omega <1$}. There  exist  some other parameter regimes in which two speeds are the same as shown  in Fig.\,{\ref{mean_field_summary}}.
 In the case of the same speeds, one can use the binary BEC systems to simulate the massive Klein-Golden equation in curved spacetime where the gapped modes exhibit the dispersion relation of a massive relativistic particle
 by regarding the sound speed as the speed of light~\cite{visser}. Here in this work, we do not restrict ourselves to these particular choices of the parameters and consider all parameter regimes with $c_{\phi} > c_{\chi}$,  $c_{\phi} < c_{\chi}$, and $c_{\phi} = c_{\chi}$ separately.

 As for the $k^4$ term of the $\phi $ mode in~(\ref{coeffs}),
its coefficient $ \zeta_\phi$ for $A/\Omega <1$ is approximated by
\begin{align}
&\zeta_\phi\simeq\no
&\frac{\gamma^2(1+\gamma)\{{[(g_{11}-g_{12})\rho_0+\Omega]^2
+\gamma[(g_{22}-g_{12})\rho_0+\Omega]^2\}}}{16m_1^2(\Omega-A)^2}\no
&+{\cal{O}}(|\Delta|),\hspace{2.95cm} {\text{for}}\quad \Delta \rightarrow 0,\no
&\zeta_\phi \simeq\,  \frac{\gamma^3}{4m_1^2}+{\cal{O}}(1/\Delta^2),\hspace{1.cm}  {\text{for}}\quad \Delta \ll 0\quad\text{or}\quad \Delta \gg 0 \, , \label{zetaphi_app_a}
\end{align}
and, for $A/\Omega>1$,
\begin{align}
&\zeta_\phi\simeq \frac{\gamma^2}{4m_1^2}+\frac{\gamma^2(1-\gamma)^2\,\Omega^2}{16m_1^2A^2}+\mathcal{O}(\vert \Delta \vert), \hspace{0.5cm} \text{for} \quad \Delta \rightarrow 0^{+},\no
&\zeta_\phi\simeq \frac{\gamma^3}{4m_1^2}-\frac{\gamma^2(1-\gamma)^2\,\Omega^2}{16m_1^2A^2}+\mathcal{O}(\vert \Delta \vert), \hspace{0.5cm} \text{for} \quad \Delta \rightarrow 0^{-}.\label{zetaphi_app_A}
\end{align}
Together with the $k^2$ term,
the dispersion relation of the gapless modes  becomes $\omega_{\phi}^2= c^2_{\phi} k^2+ (k^2/2m)^2$  of the Bogoliubov picture when $g_{11}=g_{22}$ and $m_1=m_2$ of the binary BECs, which  correspond to the binary systems with the same atoms in two different hyperfine states with the identical scattering lengths~\cite{wein}.
This can be found either  by substituting the analytical results of  mean-field solutions in the limit of $\Delta \rightarrow 0$ and $\vert \Delta \vert \gg0$ into the expression of $\zeta_\phi$ in (\ref{zetaphi_app_a}) and (\ref{zetaphi_app_A}) or by performing numerical calculations with all possible values of $\Delta$ for both $\Omega >A$ and $\Omega <A$.

Let us now consider the equal mass situation, upon which all subsequent numerical studies are based, with
$m_1=m_2=m$ and $\gamma=1$.
 We also choose the parameters $g_{11}=g_{22}$ and $g_{12}$,  given by  the system of the same atoms with different hyperfine states, for $A/\Omega <1$ ($A/\Omega >1$) with the Hamilton density shown in~Figs.\ref{mean_field_H_A}(a)--\ref{mean_field_H_A}(c) [in Figs.\,\ref{mean_field_H_A}(d)--\ref{mean_field_H_A}(f)] as changing $\Delta$.  The associated velocities of the gapless and the gapped modes are shown in Fig.\,\ref{c2} as a function of $\Delta$ for a value of $A$, consistent with the analytical results [(\ref{cphi_a}),(\ref{cchi_a}),(\ref{cphi_A}),(\ref{cchi_A})] discussed in the previous paragraph. In Fig.\,\ref{mean_field_summary}, we show the plot  in terms of the parameters ($\Delta, A$) to indicate relative values between two speeds of the gapless and gapped modes.
The parameters lying along the dotted white line of Fig.\,\ref{mean_field_summary} give the same speeds of two modes whereas $c_{\phi} > c_{\chi}$ ($c_{\chi} > c_{\phi}$)  is for the parameter regime inside (outside) the dotted white line.
Also, the graph of the effective mass  $m_{\text{eff}}$ is shown in Fig.\,\ref{meff} consistent with (\ref{meff_a}) and (\ref{meff_A}) assuming $g_{11}=g_{22}$.

In order to study the effects of interaction  between the gapless and the gapped modes, we first rewrite  $\mathcal{L}_\text{c}$ (\ref{L_c}) in terms of the $\phi$ and $\chi$ fields as
\begin{align}
\mathcal{L}_\text{c}=&\quad\frac{\lambda_{22}}{m_1\bl_0}\Big(\frac{1}{\sqrt{2}}\Big)^3(\dot{\chi}+\dot{\phi})(\nabla\chi+\nabla\phi)^2\no
&-\frac{\lambda_{11} \gamma}{m_1\bl_0}\Big(\frac{1}{\sqrt{2}}\Big)^3(\dot{\chi}-\dot{\phi})(\nabla\chi-\nabla\phi)^2\no
&-\frac{\lambda_{12} \gamma}{2m_1\bl_0}\Big(\frac{1}{\sqrt{2}}\Big)^3(\dot{\chi}+\dot{\phi})(\nabla\chi-\nabla\phi)^2\no
&+\frac{\lambda_{12}}{2m_1\bl_0}\Big(\frac{1}{\sqrt{2}}\Big)^3(\dot{\chi}-\dot{\phi})(\nabla\chi+\nabla\phi)^2 \, .
\end{align}
Among all the contributions, we prioritize the coupling terms of the form $ \chi\,\phi^2$ and ignore the other terms. These will contribute the loop effects by integrating out the $\chi$ field, presumably to be smaller when compared with the terms of our interest.
Thus, we define the interaction Lagrangian density as
\begin{align}
\mathcal{L}_\text{int}=\frac{\alpha_t}{m_1}\,\dot\chi(\nabla\phi)^2+\frac{\alpha_s}{m_1}\,\nabla\chi \,\cdot(\dot{\phi}\,\nabla{\phi})
\end{align}
with the coupling constants
\begin{align} \label{alpha}
&\alpha_t=\frac{1}{\bl_0}\Big(\frac{1}{\sqrt{2}}\Big)^3\left[\lb-\gamma\la+\frac{\lab}{2}(1-\gamma)\right]\,,\\[7pt]
&\alpha_s=\frac{1}{\bl_0}\,\Big(\frac{1}{\sqrt{2}}\Big)\left[\lb-\gamma\la-\frac{\lab}{2}(1-\gamma)\right].
\end{align}

In the binary BEC system, the gapped mode affects the dynamics of the wave propagation of the gapless $\phi$ mode along the sound cone. In other words, the intrinsic quantum stochasticity of the gapped modes makes the sound cone fluctuate. We will derive a formalism to study this effect in the next section.

\section{Langevin equation and induced stochastic sound cone metric}\label{sec3}
The whole system now consists of the $\phi$ field of the gapless mode and $\chi$ field of the gapped modes. In our approach, the $\phi$ field is our subsystem of interest and $\chi$ field is treated as the environment, which will be integrated out later.
Thus, the full action takes the form
%Now the whole system is considered to be the $\phi$ field of the gapless mode, the system of interest, and the %$\chi$ field of the gapped modes, treated as the environment to be with
%the following full action:
%
\begin{align}
S[\phi,\chi]=S_\text{sys}[\phi]+S_\text{env}[\chi]+S_{\text{int}}[\phi,\chi]\,.
\end{align}
In (\ref{S_sys}) the actions for the system and the environment are given by the terms quadratic in each field
\begin{align} \label{S_sys}
S_\text{sys}[\phi]=&\int{d^{D+1} x\left[\frac{\Lambda_\phi}{2}(\partial_t\phi)^2-\frac{\rho_{\phi}}{4m_1 }(\nabla\phi)^2\right]}\, ,\\[6pt]
S_\text{env}[\chi]=&\int{d^{D+1} x\left[\frac{\Lambda_\chi}{2}(\partial_t\chi)^2-\frac{\rho_{\chi}}{4m_1}(\nabla\chi)^2-\frac{M^2}{2} \chi^2\right]} \,, \label{S_env}
\end{align}
and the action of the interaction terms is
\begin{align}
S_\text{int}[\phi,\chi]=\int d^{D+1} x\,\left[\frac{\alpha_t}{m_1} \dot\chi(\nabla\phi)^2+\frac{\alpha_s}{m_1}\,\nabla\chi \,\cdot(\dot{\phi}\,\nabla{\phi})\right].\label{sint}
\end{align}
Turning off the interaction between the gapped and the phonon modes, the phonon field would obey the wave equation
\be \label{wave_0}
\ddot \phi (x) - c_{\phi}^2 \, \nabla^2 \phi(x)=0 \, .
\ee
In general when the coupling constants are nonzero, the coarse-graining of the degrees of freedom of the gapped $\chi$ field will influence the dynamics of the phonon $\phi$ field, introducing
the stochastic effects on the acoustic metric of the phonon via the Langevin equation.
In this nonequilibrium system, the complete information of out-of-equilibrium dynamics can be  determined by
the time-dependent density matrix  $\rho (t)$ of the  whole system.

The method, the so-called closed-time-path formalism, enables us to calculate the evolution of the density matrix that has been prepared at some particular initial time $t_i$.
 We assume that the
initial density matrix at time $t_i$ can be factorized as
\begin{equation}\label{initialcond}
    \rho(t_i)=\rho_{\phi}(t_i)\otimes\rho_{{\chi}}(t_i)\, ,
\end{equation}
where $\rho_{\phi}(t_i)$ and  $\rho_{{\chi}}(t_i)$ are
 the initial density matrix of the system and the environment.
The  full dynamics  of the whole system described by the density matrix ${ \rho}(t)$ evolves unitarily
according to
\begin{equation}
{ \rho} (t_f) = U(t_f, t_i) \, { \rho} (t_i) \, U^{-1} (t_f,
t_i )\,,
\end{equation}
with $ U(t_f,t_i) $ the time evolution operator, involving  the degrees of freedom of
the system and environment.  The reduced density matrix $\rho_r$ of the system can be obtained
by tracing over the environmental degrees of freedom in the full density matrix, thus obtaining the influence functional.
The idea here is to construct the so-called Feynman-Vernon influence functional, defined by
\begin{widetext}
\begin{align}
\mathcal{F}&[\phi^{+},\,\phi^{-}]= \exp i\, S_{\text{IF}}[\phi^{+},\phi^{-}]\no
&=\int{d\chi_f\int{d\chi_i\int{d\chi'_i}}}\,\rho_\chi(\chi_i,\chi'_i,t_i)\int^{\chi_f}_{\chi_i}{D\chi^{+}{\int^{\chi'_f}_{\chi'_i}{D\chi^{-}}}}\exp i\Big(S_{\text{env}}[\chi^{+}]+S_{\text{int}}[\phi^{+},\chi^{+}]-S_{\text{env}}[\chi^{-}]-S_{\text{int}}[\phi^{-},\chi^{-}]\Big)
\label{sinf} \, ,
\end{align}
\end{widetext}
where $\pm$ denotes the path integral forward (+) and backward ($-$) in time.
The influence functional offers the complete information on how the environment affects the system.

In nonequilibrium quantum field theory, the time evolution of the reduced density matrix of system $\phi(x)$ reads
\begin{widetext}
\begin{align}
\rho_r(\phi_f,\phi'_f,t)=\int d\phi_i\int d\phi'_i\int^{\phi_f}_{\phi_i} D\phi^{+}\int ^{\phi'_f}_{\phi'_i}D\phi^{-}\,\rho_{\phi} (\phi_i,\phi'_i,t_i)\,\exp i\Big(S_\text{sys}[\phi^{+}]-S_\text{sys}[\phi^{-}]\Big)\mathcal{F}[\phi^{+},\phi^{-}]\,.\label{den}
\end{align}
It is sufficient to consider the second order perturbation up to $\alpha^2$ in~(\ref{sinf}) with the interaction terms in~(\ref{sint}) giving
\begin{align}
S_{\text{IF}}[\phi^{+},\phi^{-}] =& \frac{i}{2}\int d^{D+1} x\int  d^{D+1} x' \no
 &\,\Big\langle \Big(\frac{\alpha_t}{m_1}\,\dot\chi^{+}(\nabla\phi^{+})^2+\frac{\alpha_s}{m_1}\nabla\chi^{+}\cdot(\dot{\phi}^{+}\nabla\phi^{+})\Big)[x]\times\Big(\frac{\alpha_t}{m_1}\,\dot\chi^{+}(\nabla\phi^{+})^2+\frac{\alpha_s}{m_1}\nabla\chi^{+}\cdot(\dot{\phi}^{+}\nabla\phi^{+})\Big)[x']\no
&-\Big(\frac{\alpha_t}{m_1}\,\dot\chi^{+}(\nabla\phi^{+})^2+\frac{\alpha_s}{m_1}\nabla\chi^{+}\cdot(\dot{\phi}^{+}\nabla\phi^{+})\Big)[x]\times\Big(\frac{\alpha_t}{m_1}\,\dot\chi^{-}(\nabla\phi^{-})^2+\frac{\alpha_s}{m_1}\nabla\chi^{-}\cdot(\dot{\phi}^{-}\nabla\phi^{-})\Big)[x']\no
&-\Big(\frac{\alpha_t}{m_1}\,\dot\chi^{-}(\nabla\phi^{-})^2+\frac{\alpha_s}{m_1}\nabla\chi^{-}\cdot(\dot{\phi}^{-}\nabla\phi^{-})\Big)[x]\times\Big(\frac{\alpha_t}{m_1}\,\dot\chi^{+}(\nabla\phi^{+})^2+\frac{\alpha_s}{m_1}\nabla\chi^{+}\cdot(\dot{\phi}^{+}\nabla\phi^{+})\Big)[x']\no
&+\Big(\frac{\alpha_t}{m_1}\,\dot\chi^{-}(\nabla\phi^{-})^2+\frac{\alpha_s}{m_1}\nabla\chi^{-}\cdot(\dot{\phi}^{-}\nabla\phi^{-})\Big)[x]\times\Big(\frac{\alpha_t}{m_1}\,\dot\chi^{-}(\nabla\phi^{-})^2+\frac{\alpha_s}{m_1}\nabla\chi^{-}\cdot(\dot{\phi}^{-}\nabla\phi^{-})\Big)[x']\Big\rangle_{\chi}\, , \label{S_if}
\end{align}
\vspace{1cm}
\end{widetext}
where $\langle ... \rangle_{\chi}$ means to take the average over the initial density matrix of the $\chi$ field in its vacuum state at zero temperature.
Integrating out the degrees of freedom of the $\chi$ field introduces the $\chi$ propagators,
\begin{align}
\langle \chi^+ (x) \chi^+ (x') \rangle=&\,\, \theta(t-t') \, \langle \chi (x)
\chi (x') \rangle \no&+ \theta(t'-t) \, \langle \chi (x')
\chi (x) \rangle  \nonumber\\
\langle \chi^- (x) \chi^- (x') \rangle =&\,\, \theta(t-t') \, \langle \chi (x') \chi
(x) \rangle \no&+ \theta(t'-t) \, \langle \chi (x)
\chi (x') \rangle \nonumber\\
\langle \chi^+ (x) \chi^- (x') \rangle =& \langle \chi (x') \chi
(x) \rangle
\nonumber\\
\langle \chi^- (x) \chi^+ (x') \rangle =& \langle \chi (x) \chi
(x') \rangle ,\label{noneq_greenfun}
\end{align}
and results in an effective nonlocal action
\be
S_{\text{eff}}[\phi^+,\phi^-]=S_{\text {sys}} [\phi^+]- S_{\text {sys}} [\phi^-]+ S_{\text{IF}}[\phi^{+},\phi^{-}] \, . \label{S_eff}
\ee

The $\chi$ propagators involve time-ordered and anti-time-ordered propagators and two Wightman propagators. All of them can be constructed by the action of the $\chi$ field in (\ref{S_env}).
 To do so, we expand the $\chi$ field in terms of creation and annihilation operators as
\begin{align}
\hat{\chi}(\mathbf{x},t)=\int \frac{d^Dk}{(2\pi)^D}\left[\hat{a}_\mathbf{k}f_k(t)e^{i\mathbf{{k}}\cdot\mathbf{x}}+\hat{a}^\dagger_\mathbf{k}f_k^\star(t)e^{-i\mathbf{{k}}\cdot\mathbf{x}}\right],
\end{align}
with the mode functions
\be
f_k(t)=\frac{1}{\sqrt{2\ob\Lambda_\chi}} \, e^{-i\ob t} \, .
 \ee
The field operators $\hat{a}_k$, $\hat{a}^\dagger_k$ obey the commutation relations:
\bea
&&[\hat{a}_\mathbf{k},\hat{a}_{\mathbf{k}'}]=0 \, , \,\,\,\, [\hat{a}^{\dagger}_\mathbf{k},\hat{a}^\dagger_{\mathbf{k}'}]=0 \, , \nonumber\\
&&[\hat{a}_\mathbf{k},\hat{a}^\dagger_{\mathbf{k}'}]=\delta^{(D)}(\mathbf{k}-\mathbf{k}').
\eea
The relevant Green's function to the Langevin equation  is   the Hadamard function $H$ given by
\begin{align} \label{H}
 H(x,x')=&2\, \langle \{ \chi(x), \chi (x') \} \rangle \nonumber\\[6pt]
 =&\,\int \frac{d^D { k}}{ ( 2\pi)^D} \frac{4  \cos[ \omega_{\chi, k} (t-t')]}{ \omega_{\chi, k} \Lambda_\chi} \, e^{-i {\bf k} \cdot ({\bf x}-{\bf x}')} \,,
\end{align}
with $\omega_{\chi, k}=\sqrt{c_{\chi}^2 k^2+ m_{\text{eff}}^2}$.
To recover the semiclassical dynamics of the $\phi$ field~(\ref{wave_0}), we then introduce  the ``center of the mass" field $\phi$ and the ``relative" field $R$ as
\begin{flalign}
\phi(x)=\frac{1}{2}(\phi^+(x)+\phi^-(x)) \, ,\qquad  R(x)=\phi^+(x)-\phi^-(x),
\end{flalign}
and  consider the fluctuation field $R$ about the field $\phi$.
 In deriving the  Langevin equation, we only retain the terms in (\ref{S_eff}) linear and quadratic in $R$ assuming that quantum fluctuations of the $\chi$ modes just give perturbative effects to the wave propagation. Additionally, we consider the sound wave with small amplitude so that the terms linear in $\phi$ are considered.
 The action of the system in (\ref{S_sys}) can be rewritten with $\phi$ and $R$ fields as
\begin{align} \label{S_sys_R}
&S_{\text{sys}}[\phi^+=\phi+{R}/{2}]-S_{\text{sys}}[\phi^-=\phi-{R}/{2}] \no[7pt]
&\hspace{0.5cm}=\int{d^{D+1} x \; R(x)\Big(-\Lambda_\phi\partial_t^2 \phi(x) +\frac{\rho_{\phi}}{2m_1}\nabla^2 \phi(x)\Big)} \,.
\end{align}
As for the action  from the influence function in~(\ref{S_if}), $S_{\text{IF}}$ splits into two types of the terms, linear and quadratic in $R$ respectively. The terms linear in $R$, the backreaction effects to give the modifications on the wave propagation from the gapped modes, turn out to involve nonlinear terms in $\phi$, which can be ignored as long as the field $\phi$ remains small. Quadratic terms in $R$ in $S_{\text{IF}}$  are then linearized by the introduction of the noise $\xi$. So, together with (\ref{S_sys_R}), $S_{\text{eff}}[\phi^+=\phi+R/2,\phi^-=\phi-R/2]$ takes the form
\begin{widetext}
\begin{align}
S_\text{eff}[\phi(x),R(x),\xi (x) ]=&\int\;d^{D+1} x\bigg[\ \Big(-\Lambda_\phi{\ddot \phi(x)} +\frac{\rho_{\phi}}{2m_1}\nabla^2 \phi(x)\Big)R(x)\,+\frac{\alpha_t}{m_1} \,  \dot{\xi}(x) \, \nabla \phi(x)\cdot\nabla R(x)\no
&\hspace{3cm}+\frac{\alpha_s}{2 m_1}\dot{\phi}(x)\, \nabla\xi(x)\cdot\nabla R(x)+\frac{\alpha_s}{2 m_1} \dot{R}(x)\, \nabla \phi(x)\cdot\nabla \xi(x))
\bigg]\,,
\end{align}
\vspace{0.5cm}
\end{widetext}
in terms of  the Gaussian noise $\xi$ with the distribution function given by the Hadamard function (\ref{H})
\be \label{noise_H}
\langle \xi (x) \xi (x') \rangle= H (x,x') \, .
\ee
In the semiclassical approximation,
 %by considering  sound waves,
 the wave equation that  describes the propagation of long wavelength phonons in a stochastic background is derived by $
{\delta S_\text{eff}[\phi (x) ,R (x) ,\xi (x)]}/{\delta R(x)}=0 $, leading to the Langevin equation
\begin{widetext}
\bea
&&\ddot{\phi}(x)-\frac{\rho_{\phi}}{2m_1\Lambda_{\phi}}\Big(1-\frac{2\alpha_t}{\rho_{\phi}}\dot{\xi}(x)\Big)\nabla^2\phi(x)=-\frac{1}{\Lambda_\phi}\left[\frac{\alpha_t}{m_1} \nabla\dot{\xi}(x)\cdot\nabla \phi(x)+\frac{\alpha_s}{2 m_1} \nabla\cdot\left(\dot{\phi}(x)\nabla\xi(x)\right)+\frac{\alpha_s}{2 m_1}\frac{\partial}{\partial t}\Big(\xi(x)\nabla \phi(x)\Big)\right].\no
\eea
\end{widetext}
As a result, we can interpret $c_{\xi}$,
\begin{align}
c_{\phi,\xi}^2&=c_{\phi}^2\Big(1-\frac{2 \alpha_t}{\rho_{\phi}}\dot{\xi}(x)\Big)\,,
\end{align}
as a stochastic speed of sound in the long wavelength limit.
The multiplicative noise term  $\dot\xi \nabla^2 \phi$ comes from the coupling terms $G(\delta\theta_1) G(\delta\theta_2)$ between  the $\theta_1$ and $\theta_2$ fields in terms of the Galilean invariants.
The underlying Galilean symmetry of the  system provides the type of the coupling to the gapless modes from the gapped modes.
Additionally, the noise $\xi$ essentially is induced from the fluctuation field $\chi$ with the known field correlation function given by the Hadamard function~(\ref{noise_H}).
Our main interest is to explore the effect of the fluctuation background on the propagation of the photon with the trajectory determined by its operator-valued acoustic metric
\be
c_{\phi}^2 dt^2-\Big(1-\frac{2 \alpha_t}{\rho_{\phi}}\dot{\xi}(x)\Big) d{\bf x}^2=0 \, .
\ee
This effect can be tested by the TOF variation of the phonons or sound waves between a source and a detector to be discussed in the next section. The experiments that we have in mind are along the lines of~\cite{ham,and} in which the sound speed of the density wave is measured in either one- or two-components BEC systems.

\section{Time-Of-Flight variance}\label{sec4}

The phonon propagates along the sound cone,
which is  given by the null geodesic
\begin{align}
c_{\phi}^2dt^2=d\mathbf{x}^2+h_{ij}dx^idx^j,
\end{align}
where $h_{ij}=(2\alpha_t/\rho_\phi) \dot{\xi}(x)\delta_{ij}$.
Consider a wave packet of the sound that propagates a spatial distance $r$; then the flight time can be estimated as
\begin{align}
T =\int_0^Tdt\approx\int_0^{r_0} dr \, \frac{1}{c_{\phi}} \bigg[ \, 1+\frac{1}{2} h_{ij} { n}^i { n}^j \, \bigg] \, ,
\end{align}
where $dr=d \vert {\bf x }\vert$, and ${\bf n}^i=d{\bf x}^i/dr$ is a unit vector along the propagation of the sound wave.  The initial time $t_i$ is set to be $t_i=0$. The local velocity $c_{\phi}$ is evaluated on the unperturbed path of the wave $ r(t)$, such that we take it along the $z$-direction with a fixed distance $z_0$. With the vanishing  stochastic average of  $h_{ij}$, $\langle h_{ij} \rangle=0$, the stochastic average of the flight time  is $
\langle T \rangle =z_0/c_{\phi}. $
Nevertheless, the influence of quantum fluctuations of the gapped modes induces the variation of the travel time in terms of the noise $\xi$, which is  given by
\begin{align}
&(\Delta T)^2=\langle T^2 \rangle-\langle T\rangle^2\no
&= \frac{1}{4}\int_0^T dt \int_0^T dt' \,\, { n}^i   { n}^j  { n}^l  { n}^m
\langle h_{ij} (r(t), t) h_{lm} (r'(t'),t') \rangle  \no
&=\frac{ \alpha_t^2}{ \rho_{\phi}^2}\int_0^Tdt\int_0^Tdt'\langle\dot{\xi}(x(t),t)\dot{\xi}(x'(t'),t')\rangle \, .
\end{align}
Using the correlation function in~(\ref{H}),  the flight time variance becomes
\begin{align}
(\Delta T)^2&=\frac{ \alpha_t^2}{ \rho_{\phi}^2} \int\frac{d^D k}{(2\pi)^D} \int_{0}^{T}dt\int_{0}^{T}dt'\times\no
&\frac{2 }{\omega_{\chi,k}\Lambda_{\chi} } \frac{\partial}{\partial t} \frac{\partial}{\partial t'} \Big\{ \, \cos[ \omega_{\chi, k} (t-t')] \, e^{-i {\bf k} \cdot ({\bf x}-{\bf x}')} \Big\} \, .
\label{nosw}
\end{align}
Notice that this variation of the flight time can be experimentally tested by repeatedly measuring the flight time traveled by the sound waves for a fixed distance with identical wave packet forms.
 It is seen that the integral over the momentum is found to have UV divergence for $ D\geq 1$. Here we introduce a suitable switching function, giving a smooth switching process on the Rabi coupling, to regularize the momentum integral, as
\begin{align}
s(u)=\frac{1}{\pi }\left[\arctan\left(\frac{u}{\tau}\right)+\arctan\left(\frac{T-u}{\tau}\right)\right]\, .
\end{align}
The parameter $\tau$ is the timescale for turning on/off the coupling process.
With the switching function suitably substituted into the time integral, the flight time variance can be expressed as
\begin{align}
&(\Delta T)^2=&\no
&\frac{\alpha_t^2}{ \rho_{\phi}^2 } \frac{2}{\Lambda_{\chi}} \int\frac{d^D k}{(2\pi)^D}  \frac{1}{\omega_{\chi,k}} \,\Bigg\vert {\int_{-\infty}^{\infty} du\, s(u)\frac{\partial}{\partial u}e^{-i(\ob -\ca k\cos\theta)u}}\Bigg\vert^2 \no[6pt]
&= \frac{   \alpha_t^2}{ \rho_{\phi}^2 } \frac{2}{\Lambda_{\chi}} \int\frac{d^D k}{(2\pi)^D} \, g(\ob -\ca k\cos\theta)\,. \label{gpm}
\end{align}
The function $g(\ob -\ca k\cos\theta)$ in the integrand is defined as
\begin{align}
&g(\omega_{\chi,k} - c_{\phi} k\cos\theta)\no
&=\frac{1}{\omega_{\chi,k} }\Bigg\vert{\int_{-\infty}^{\infty}du\,s(u)\frac{\partial}{\partial u}e^{-i(\omega_{\chi,k} - c_{\phi} k\cos\theta)u}}\Bigg\vert^2\no
&=\frac{4}{\omega_{\chi,k} }e^{-2\tau\vert {\omega_{\chi,k} - c_{\phi} k\cos\theta}\vert}\sin^2{\bigg[\frac{T}{2}(\omega_{\chi,k} - c_{\phi} k\cos\theta)\bigg]}\, .\label{gpm2}
\end{align}
For $D \geq 2$, the momentum integration, even after introducing the switching function, still suffers UV divergence. In these cases, one may cut off the integral at some particular $k_{\Lambda}$.
 Here we will focus on the pseudo-one-dimensional problem. The flight time variance in one dimension with $\theta=0$ can be written from~(\ref{gpm}) and (\ref{gpm2})  as
\begin{align}
&(\Delta T)_{\text{1d}}^2=\no
&\frac{  \alpha_t^2}{ \rho_{\phi}^2} \frac{2}{\Lambda_{\chi}}\int_{-\infty}^{\infty}\frac{dk}{2\pi}\;\frac{4}{\ob }e^{-2\tau\vert{\ob -\ca k}\vert}\sin^2{\bigg[\frac{T}{2}(\ob- \ca k)\bigg]}.\label{ft2}
\end{align}
Given the explicit expression of TOF variation, we now consider its effect.
Recall the dispersion relation of the $\chi$ field, $\omega_{\chi,k}=(c_{\chi}^2k^2+m_{\text{eff}}^2)^{1/2}$.
To proceed further with the analytical studies, we approximate  the phase
%$(\omega_{\chi,k}- c_{\phi}k)$
as follows:
\begin{align}\label{phase_app}
&\omega_{\chi,k} - c_{\phi}k=\no
&\begin{cases}
c_{\chi} k -c_{\phi} k+ \dfrac{m^2_\text{eff}}{2\, \cb k},& \text{when}\quad \dfrac{m_{\text{eff}}}{c_{\chi}} <k< \infty,\\[12pt]
\omega_{\chi,k}- c_{\phi}k ,&\text{when}\quad -\dfrac{m_{\text{eff}}}{c_{\chi}}<k<\dfrac{m_{\text{eff}}}{c_{\chi}},\\[9pt]
-\,c_{\chi} k - c_{\phi}k -\dfrac{m^2_\text{eff}}{2\, \cb k}, & \text{when}\quad -\infty < k<-\dfrac{m_{\text{eff}}}{c_{\chi}}.
\end{cases}
\end{align}
Accordingly, the integral in (\ref{ft2}) will be performed separately in the above three ranges of $k$.
%In below, the whole integral over the momentum $k$ for obtaining its analytic expression.
In the following subsections, the calculations of TOF variation will be considered separately for the cases,
$ c_{\phi} =c_{\chi}$, $c_{\phi} > c_{\chi}$ and $c_{\phi} < c_{\chi}$.
These approximate results of $(\Delta T)^2$ at large flight time $T$ will be compared later with the full numerical calculations.
From now on we restrict ourselves to the problem in one dimension, and omit the subscript $1d$ for the simplicity of notation.

\subsection{$c_{\phi}=c_{\chi}$}
Let us start the case of equal velocity.
With the help of ~(\ref{phase_app}), we can compute easily the contributions of $-\infty<k<- m_{\text{eff}}/c_{\chi}$ to the TOF variation for large  $T$, denoted with $(\Delta T)^2_{(1)}$, as
%In the region of $-\infty<k<- m_{\text{eff}}/c_{\chi}$ with the phase approximation above, at the contribution to %the TOF variation can be obtained as
%
\begin{align}
&(\Delta T)^2_{(1)} (T \rightarrow \infty)\no &\simeq\frac{{\alpha_t}^2}{\rho_{\phi}^2} \frac{1}{\pi \Lambda_{\chi}} \int_{-\infty}^{-m_{\text{eff}}/c_{\chi}}dk\frac{4}{(- c_{\chi}k)}\,e^{4\tau c_{\phi}k}\,\sin^2[\,T c_{\chi}k]\no
&\simeq -\frac{\alpha_t^2}{\rho_{\phi}^2}\frac{2}{\pi c_{\chi}\Lambda_\chi}\text{Ei}(-4\tau \,m_\text{eff}) \no
&\simeq   \frac{\alpha_t^2}{\rho_{\phi}^2}\frac{2}{\pi c_{\chi}\Lambda_\chi} \bigg[  -\gamma -\ln( 4 \tau m_{\text{eff}})+ 4\tau m_{\text{eff}} +\mathcal{O}(\tau^2 m_\text{eff}^2)\bigg]\, .\label{ft1_1}
\end{align}
In (\ref{ft1_1}), we have approximated
$\sin^2[T c_{\chi}k] \approx 1/2$
%-\cos[2T c_{\chi}k]/2\approx 1/2$
due to its rapidly oscillatory behavior in high momentum $k$,
so that $(\Delta T)^2_{(1)}$ can be simply written as  the exponential integral function, $\text{Ei}$,  with no explicit $T$ dependence.
We also assume that $\tau m_\text{eff}\ll 1 $, meaning that   the interaction is
quickly switched on/off within the timescale $\tau$ for expectedly having  large $(\Delta T)^2 $ during the flight time $T$.

In the interval of momentum $- m_{\text{eff}}/c_{\chi}<k< m_{\text{eff}}/c_{\chi}$, it is straightforward to estimate the integral as
\begin{widetext}
\begin{align}
 (\Delta T)^2_{ (2)} (T\rightarrow \infty) &\simeq
 \frac{{\alpha_t}^2}{\rho_{\phi}^2} \frac{1}{\pi \Lambda_{\chi}}  \int_{-m_{\text{eff}}/c_{\chi}}^{m_{\text{eff}}/c_{\chi}}dk\frac{4}{\omega_{\chi,k}}\,e^{-2\tau|\omega_{\chi,k}- c_{\phi}k|}\sin^2\bigg[\,\frac{T}{2}(\omega_{\chi,k}- c_{\phi}k)\bigg] \no
&\simeq \frac{{\alpha_t}^2}{\rho_{\phi}^2} \frac{1}{\pi \Lambda_{\chi}} \int_{-m_{\text{eff}}/c_{\chi}}^{m_{\text{eff}}/c_{\chi}}dk\frac{4}{2\sqrt{c_{\chi}^2k^2+m_{\text{eff}}^2}}\left[1-2\tau \big(\sqrt{c_{\chi}^2k^2+m_{\text{eff}}^2}-c_{\chi}k\big)\right]\no
& \simeq \frac{{\alpha_t}^2}{\rho_{\phi}^2} \frac{2\,}{\pi\Lambda_\chi c_{\chi}}\bigg[\ln{\bigg(\frac{\sqrt{2}+1}{\sqrt{2}-1}\bigg)-4\tau m_\text{eff}+\mathcal{O}(\tau^2 m_\text{eff}^2)}\bigg],
\label{ft1_2}
\end{align}
\end{widetext}
where we have approximated the square of the sine function by $1/2$ and expanded the exponential  term to linear order in $\tau$.

 %the square of the sine function has been  approximately by $1/2$
 %for such a rapidly oscillating function when $T$ is large, and for $\tau m_\text{eff}\ll 1 $, is  to expand

The contribution to $(\Delta T)^2$ from the interval of the momentum $m_{\rm eff}/c_{\chi}<k<\infty$ reads
%
% Finally,  we consider the momentum integral over the region of $m_{\rm eff}/c_{\chi}<k<\infty$, together with the above phase approximation~(\ref{phase_app}),  given by
 %
 \begin{align}
&(\Delta T)^2_{ (3)} (T\rightarrow \infty)\no
&\simeq  \frac{{\alpha_t}^2}{\rho_{\phi}^2} \frac{1}{\pi \Lambda_{\chi}}   \int_{m_{\text{eff}}/c_{\chi}}^{\infty}dk\, \frac{4}{ c_{\phi}k}e^{-\tau m^2_\text{eff}/ c_{\chi}k}\sin^2\bigg[\frac{T m^2_\text{eff}}{4 c_{\chi}k}\bigg] \, .
\end{align}
Notice that for large $T$ and $k$,
their ratio can be small and  may
make the square of the sine function less oscillatory when $k$ varies,
leading to a
%end result in this regime of the $k$-integral
growth of the TOF fluctuations in $T$. To see this,
we change the variable to  $k=1/u$, and the integral  becomes
\begin{widetext}
\begin{align}
(\Delta T)^2_{(3)} (T \rightarrow \infty )&\simeq  \frac{{\alpha_t}^2}{\rho_{\phi}^2} \frac{1}{\pi \Lambda_{\chi}}  \int_{0}^{1/(m_{\text{eff}}/c_{\chi})}du\frac{4}{ c_{\chi}u}\,e^{-\tau m^2_\text{eff}u / c_{\chi}} \,\sin^2\bigg[\frac{T m^2_\text{eff}}{4c_{\chi}}u\bigg] \no[6pt]
&\simeq  \frac{{\alpha_t}^2}{\rho_{\phi}^2} \frac{4}{\pi \Lambda_{\chi}}    \bigg[
\int_{0}^{4\pi c_{\chi}/Tm_\text{eff}^2}du\frac{1}{ c_{\chi}u}\,\sin^2\bigg[\frac{T m^2_\text{eff}}{4c_{\chi}}u\bigg]+\int_{4\pi c_{\chi}/Tm_\text{eff}^2}^{1/(m_{\text{eff}}/c_{\chi})}du\frac{1}{2 c_{\chi}u}
\bigg]\no[6pt]
&\simeq \frac{{\alpha_t}^2}{\rho_{\phi}^2} \frac{2}{\pi \Lambda_{\chi} c_{\chi} } \Big[\gamma+
\ln(2\pi)-Ci(2\pi)+\ln(Tm_{\text{eff}}/ 4 \pi)+\mathcal{O}(\tau^2m_\text{eff}^2)
\Big]\,.\label{ft1_3}
\end{align}
\end{widetext}
The integration over $u$ can be carried out in the regions of $0< u < 4\pi c_{\chi}/Tm_\text{eff}^2$ with the growth of the TOF variation in $T$, and also in the regime of $4\pi c_{\chi}/Tm_\text{eff}^2<u< 1/(m_{\text{eff}}/c_{\chi})$, where the square of the sine function can be safely approximated by  $1/2$ for large $T$, leading to a $T$-independent saturated value.
Adding up the contributions from the three momentum integrals, (\ref{ft1_1}), (\ref{ft1_2}) and (\ref{ft1_3}), we arrive at the following expression for the total effect of the flight time variance:
%Thus, ends up with  the sum of the above results, and shows the growth as
\begin{align} \label{DT2_t1}
&(\Delta T)^2_{(c_{\phi}=c_{\chi})} (T \rightarrow \infty)\no[5pt]
&=(\Delta T)^2_{ (1)} +(\Delta T)^2_{ (2)} +(\Delta T)^2_{ (3)} \no[5pt]
&\simeq \frac{{\alpha_t}^2}{\rho_{\phi}^2}\frac{2}{\pi \Lambda_{\chi} c_{\chi} }\bigg[\ln{\left(\frac{\sqrt{2}+1}{\sqrt{2}-1}\cdot\frac{T}{8\tau}\right)}-Ci(2\pi)+\mathcal{O}(\tau^2m_\text{eff}^2)\bigg].
\end{align}
In this case of equal speeds, the results show the logarithmic growth of TOF fluctuation with the flight time $T$.
Of course, the conditions of equal velocities of two modes  need some  particular choices of the effective parameters for $A$ and $\Delta$ indicated  in Fig.\,\ref{mean_field_summary}.
%
%To reach the situation with equal velocities of two modes  needs some  particular choices of the  parameters for $A$ and $\Delta$ indicated  in Fig.\,\ref{mean_field_summary} so as to  exhibit the logarithmic growth of TOF fluctuations in $T$ in this system.
Whether or not the $(\Delta T)^2$ effect can grow large enough to be seen with  current experimental technologies depends on the parameters of relevance to these experiments. We will return to this point later in this section.

\subsection{$c_{\phi}>c_{\chi}$}

Following a similar analysis, we turn to the discussions of the case  $c_{\phi}>c_{\chi}$. Here the approximation $\sin^2[{T}(\ob- \ca k)/2] \rightarrow 1/2$ for large $T$ can be applied in all ranges of $k$ in the momentum integrals.  As a result,  $(\Delta T)^2 (T \rightarrow \infty)$ will saturate into the $T$-independent value.

 For the momentum interval $-\infty<k<-m_{\text{eff}}/c_{\chi}$, the contribution to $(\Delta T)^2_{ (1)}$ at large $T$ is given by
\begin{widetext}
\begin{align} \label{DT2_21}
(\Delta T)^2_{ (1)} (T \rightarrow \infty)
&=\frac{{\alpha_t}^2}{\rho_{\phi}^2} \frac{1}{\pi \Lambda_{\chi}} \int_{-\infty}^{-m_{\text{eff}}/c_{\chi}}dk\frac{4}{(- \cb k)}\,e^{2\tau (c_{\phi}+c_{\chi}) k}\,\no
&\simeq - \frac{\alpha_t^2}{\rho_{\phi}^2}\frac{2}{\pi c_{\chi}\Lambda_\chi}\text{Ei} {\left(-2\tau (c_{\phi}+c_{\chi}) m_\text{eff}/\cb\right)}\no[6pt]
&\simeq \frac{\alpha_t^2}{\rho_{\phi}^2}\frac{2}{\pi c_{\chi}\Lambda_\chi} \bigg[-\gamma-\ln{(2\tau(c_\phi+c_\chi)m_\text{eff}/c_\chi)}+2\tau(c_\phi+c_\chi)m_\text{eff}/c_\chi+\mathcal{O}(\tau^2m_\text{eff}^2)\bigg]
\, ,
\end{align}
with $\tau m_{\text{eff}} \ll 1$ as before.
Similarly, for the region $m_{\text{eff}}/c_{\chi}<k<\infty$, the integral can be estimated to be
\begin{align} \label{DT2_23}
(\Delta T)^2_{ (3)} (T\rightarrow \infty )& \simeq -  \frac{\alpha_t^2}{\rho_{\phi}^2}\frac{2}{\pi c_{\chi}\Lambda_\chi}\text{Ei} (-2\tau (c_\phi-c_\chi)  m_\text{eff}/c_\chi) \no
& \simeq \frac{\alpha_t^2}{\rho_{\phi}^2}\frac{2}{\pi c_{\chi}\Lambda_\chi}\bigg[-\gamma-\ln{(2\tau(c_\phi-c_\chi)m_\text{eff}/c_\chi)}+2\tau(c_\phi-c_\chi)m_\text{eff}/c_\chi+\mathcal{O}(\tau^2m_\text{eff}^2)\bigg]
\, .
\end{align}
Finally, in the region $-m_{\text{eff}}/c_{\chi}<k<m_{\text{eff}}/c_{\chi}$, we have
\begin{align}
(\Delta T)^2_{ (2)} (T \rightarrow \infty )&=\frac{{\alpha_t}^2}{\rho_{\phi}^2} \frac{1}{\pi \Lambda_{\chi}} \int_{-m_\text{eff}/c_\chi}^{m_\text{eff}/c_\chi}dk\frac{4}{\omega_{\chi,k}}\,e^{-2\tau|\omega_{\chi,k}- c_{\phi}k|}\no[6pt]
&\simeq \frac{{\alpha_t}^2}{\rho_{\phi}^2} \frac{1}{\pi \Lambda_{\chi}} \int_{-m_\text{eff}/c_\chi}^{m_\text{eff}/c_\chi}dk\frac{4}{\omega_{\chi,k}}\,e^{-2\tau|\omega_{\chi,k}- c_{\phi}k|}\, .\label{ft2_2}
\end{align}
%which, for  $\tau m_{\text{eff}} \ll1$ and with the  large flight time $T$,  can be further approximated by
For the conditions $\tau m_{\text{eff}} \ll1$ and large flight time $T$, (\ref{ft2_2}) can be further approximated by
\begin{align} \label{DT2_22}
(\Delta T)^2_{ (2)} (T \rightarrow \infty)& \approx \frac{{\alpha_t}^2}{\rho_{\phi}^2} \frac{2\,}{\pi\Lambda_\chi c_{\chi}}\Bigg[\ln{\left(\frac{\sqrt{2}+1}{\sqrt{2}-1}\right)-4\tau m_\text{eff}+\mathcal{O}(\tau^2 m_\text{eff}^2})\Bigg].
\end{align}
In summary, the flight time variance for the case $\ca>\cb$ is found as
\begin{align}
&(\Delta T)^2_{(c_{\phi} > c_{\chi})} (T\rightarrow \infty)=(\Delta T)^2_{ (1)} +(\Delta T)^2_{ (2)} +(\Delta T)^2_{ (3)} \no
&= \frac{{\alpha_t}^2}{\rho_{\phi}^2} \frac{2\,}{\pi\Lambda_\chi c_{\chi}} \Bigg[-2\gamma+4\tau (c_\phi/c_\chi-1) m_\text{eff}+\ln(4\tau^2 (c_\phi^2-c_\chi^2) m_\text{eff}^2/\cb^2)+\ln\left(\frac{\sqrt{2}+1}{\sqrt{2}-1}\right)+\mathcal{O}(\tau^2 m_\text{eff}^2)\Bigg],
\label{DT2_t2}
\end{align}
\end{widetext}
the saturated value in the late time limit.

\subsection{$c_{\phi}<c_{\chi}$}

%Contrary to the case of $c_{\phi} > c_{\chi}$, here we reexamine the integral over $k$ for $c_{\phi}<c_{\chi}$ in stead.
In this case, the phase $(\omega_{\chi,k}- c_{\phi}k)  \simeq  c_{\chi} k -c_{\phi} k+ ({m^2_\text{eff}}/{2\, \cb k})$ in the large $k$ limit has the  minimum value
%for $c_{\phi}<c_{\chi}$
at $k=k_m=m_\text{eff}\,c_\phi/(c_\chi\sqrt{c_\chi^2-c_\phi^2})$.
Apart from the saturated value of $(\Delta T)^2_{(c_{\phi} > c_{\chi})} (T\rightarrow \infty)$, which is given by~(\ref{DT2_t2}) by appropriate replacement of $c_{\phi} \leftrightarrow c_{\chi}$, it will be seen that the saturated value is reached in an oscillatory manner during the flight time $T$.
To proceed with the calculation of TOF fluctuations, denoted by $(\Delta T)^2_\text{osc}$, we expand the phase around $k_m$,
\begin{align}
&(\omega_{\chi,k}-c_\phi k)\approx \no
&\quad {m_\text{eff}\sqrt{c_\chi^2-c_\phi^2}}/{c_\chi}+\beta \left(k-k_m\right)^2 +{\cal{O}} (\left(k-k_m\right)^4),
\label{ph}
\end{align}
with $\beta ={(c_\chi^2-c_\phi^2)^{3/2}}/{2m_\text{eff}\,c_\chi}$.
The oscillatory behavior is encoded in the form
\begin{widetext}
\begin{align}
& (\Delta T)^2_\text{osc} (T \rightarrow \infty)=-\frac{\alpha_t^2}{\rho_\phi^2}\frac{1}{\pi \Lambda_\chi c_\chi}\int_{-\infty}^{\infty}dk\frac{4}{2c_\chi k}\, e^{-2\tau|\omega_{\chi,k}- c_{\phi}k|} \,\cos{[T(\omega_{\chi,k} -  c_{\phi}k)]} \no
&\quad\quad\simeq -\frac{\alpha_t^2}{\rho_\phi^2}\frac{2}{\pi \Lambda_\chi c_\chi} \int_{k_m-\sqrt{\pi/T \beta}}^{k_m+\sqrt{\pi/T \beta}} dk\frac{1}{k} \, \cos[T({m_\text{eff}\sqrt{c_\chi^2-c_\phi^2}}/{c_\chi}+\beta (k-k_m)^2)] \Big(1+{\cal{O}} (\tau m_{\text{eff}} ) \Big)  \no
&\quad\quad\simeq \frac{\alpha_t^2}{\rho_\phi^2}\frac{2}{\pi \Lambda_\chi c_{\chi} } \frac{c_\chi^{3/2}}{ ({c_\chi^2-c_\phi^2})^{1/4}c_\phi} \sqrt{\frac{4\pi}{T m_\text{eff} }}\times\no
&\qquad\quad\quad\quad\quad \bigg[C(\sqrt{2}) \cos{[Tm_\text{eff} \sqrt{c_\chi^2-c_\phi^2}/c_\chi ]} -S(\sqrt{2}) \sin{[Tm_\text{eff}  \sqrt{c_\chi^2-c_\phi^2}/c_\chi ]} \bigg]\Big(1+{\cal{O}} (\tau m_{\text{eff}}) \Big) \,,\label{ft3}
\end{align}
\end{widetext}
{where $\tau m_{\text{eff}}  \ll 1$ is assumed} and the momentum integral is evaluated only near $k=k_m$.
In (\ref{ft3}), the Fresnel integral functions are defined as
\begin{align*}
S(x)=\int_0^x\sin\left(\pi t^2/2\right)dt,
\end{align*}
and
\begin{align*}
C(x)=\int_0^x\cos\left(\pi t^2/2\right)dt \, .
\end{align*}
Thus, the flight time variance for $c_\chi>c_\phi$ becomes finally
\begin{widetext}
\begin{align}
(\Delta T)^2_{(c_{\phi} < c_{\chi})}&(T\rightarrow \infty)\no
&=\frac{\alpha_t^2}{\rho_\phi^2}\frac{2}{\pi \Lambda_\chi c_\chi}\Bigg\{
 -2\gamma +4\tau(c_\chi/c_\phi-1)m_\text{eff}+\ln{(4\tau^2(c_\chi^2-c_\phi^2)m_\text{eff}^2/c_\chi^2)}+\ln{\left(\frac{\sqrt{2}+1}{\sqrt{2}-1}\right)}\no[6pt]
&
\hspace{2.5cm} +\frac{c_\chi^{3/2}}{({c_\chi^2-c_\phi^2})^{1/4}c_\phi}\sqrt{\frac{4\pi}{{T m_\text{eff}}}}\bigg[C(\sqrt{2}) {\cos{\left[Tm_\text{eff}\sqrt{c_\chi^2-c_\phi^2}/c_\chi\right]}} \no[6pt]
& \hspace{2.5cm}-S(\sqrt{2}) {\sin{\left[Tm_\text{eff}\sqrt{c_\chi^2-c_\phi^2}/c_\chi\right]}}
\bigg]\bigg(1+{\cal{O}} (\tau m_{\text{eff}}  )\bigg) +\mathcal{O}(\tau^2 m_\text{eff}^2)
\Bigg\} \, ,
\label{DT2_t3}
\end{align}
\end{widetext}
which reaches the saturation in an oscillatory way for large flight time $T$.
The analytical expressions of $(\Delta T)^2 (T\rightarrow \infty)$ in each of the cases are one of  the main results in this paper, and later these are compared with the full numerical calculations using the parameters relevant to the experiments.

\begin{figure}[th]
\begin{center}
\includegraphics[width=1\linewidth, trim={1.6cm 0 2cm 0},clip]{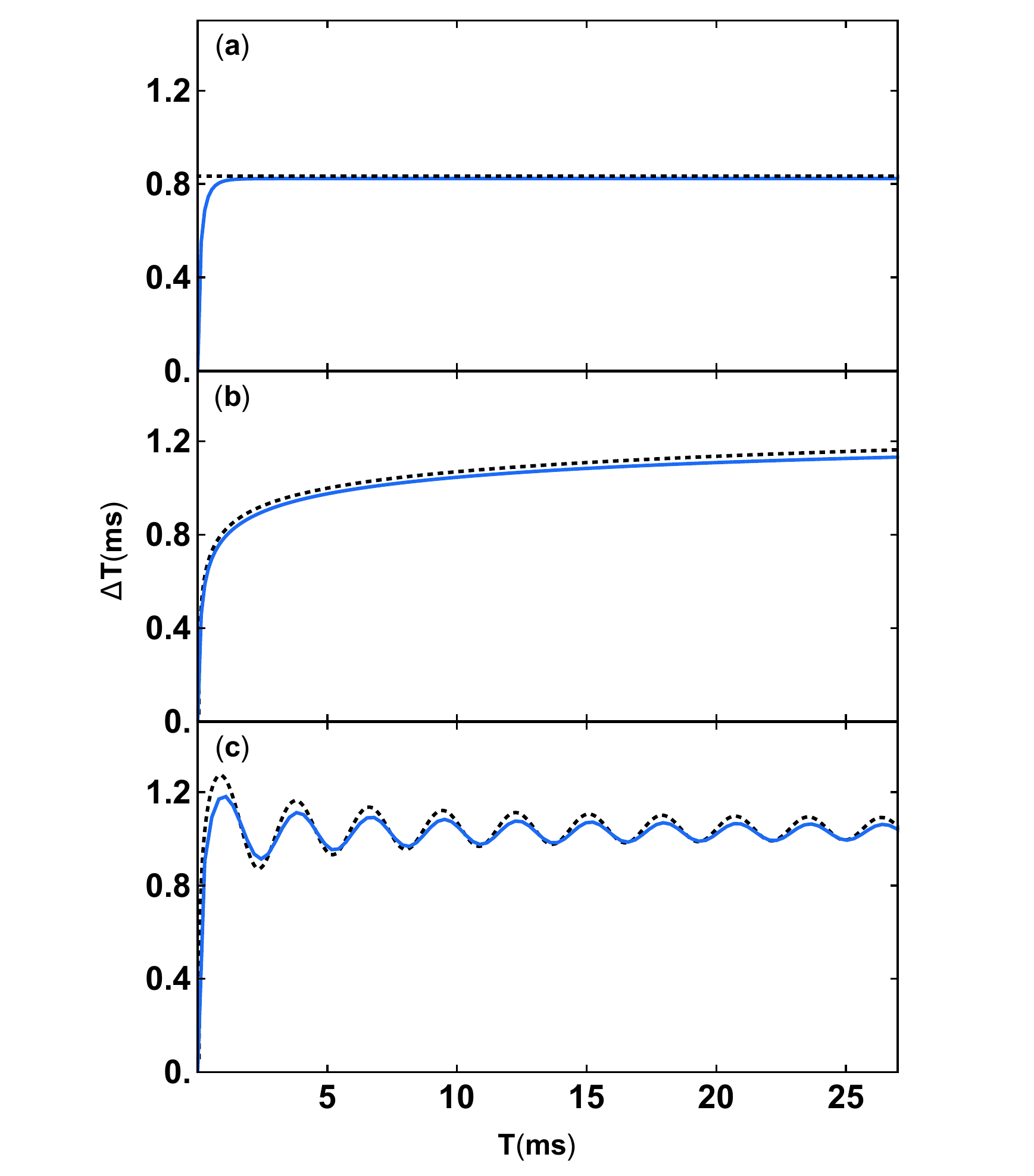}
\caption{The TOF variation $\Delta T $  as a function of the flight time $T$ for the full numerical results (solid lines) and analytical approximations~(\ref{DT2_t1}), (\ref{DT2_t2}), (\ref{DT2_t3}) (dashed lines). The  parameters are as follow: Rabi frequency $\Omega_0=200\text{Hz}$, scattering lengths $a_{11}=a_{22}=100\text{a}_0$,
 detuning $\Delta=0.0\Omega$, with different values of interspecies scattering lengths  $a_{12}$ with
(a) $a_{12}=280\text{a}_0$ ($A=43.8\Omega_0$),
 (b) $a_{12}=300\,\text{a}_0$ ($A=48.6\Omega_0$), and (c) $a_{12}=320\text{a}_0$ ($A=53.5\Omega_0$)}
\label{DT2_n2}
\end{center}
\end{figure}

To be concrete, consider a cold $^{87}\text{Rb}$ ($ m \simeq 144 \times 10^{-27} \text{kg}$)  condensate in  two hyperfine states $\vert 1\rangle=\vert F=1, m_F=1 \rangle$
and $\vert 2 \rangle=\vert F=2, m_F=-1\rangle$~\cite{zib}.  The  scattering lengths are  $a_{11}= a_{22}=100 a_0=5.3 \text{nm}$ with $a_0$ the Bohr radius.
The existence of the Feshbach
resonance at $9.1 \text{G}$ permits us to tune the  scattering length $a_{12}$ to be specified in each case later~\cite{kem}.
Let us consider two-component elongated BECs in quasi-1D geometries~\cite{det,ham} where the sizes of the condensates are about $L_z=100\mu \text{m}$  along the axial direction  and $L_r$ of order $\mu \text{m}$ in the radial direction with an aspect ratio $L_z/L_r \approx 200$~\cite{det,sab}.
We take the number of atoms $N=10000$~\cite{zib} with the number density
$\rho_0
\approx 1 \times 10^{8} \text{m}^{-1}$~\cite{det,sab,lel}. Note that in a quasi-one dimensional system, the coupling constant $g_{ij}$ can be given by the scattering length $a_{ij}$ as $g_{ij}=2 \pi \hbar^2 a_{ij}/L_r^2$ for $i,j=1,2$.
 We also choose the Rabi frequency $\Omega_0 =200 \text{Hz}$~\cite{zib} for reference with an input energy about $E=\hbar \Omega \approx 10^{-32} \text{J}$, which is much smaller than  the required thermal energy to melt down the condensates with the critical temperature $T_c \approx 10^{-29} \text{J}$, estimated from the 3D number density given by the above-mentioned parameters. In addition, the detuning parameter is set to zero ($\delta=0 \Omega$).  Also, for $m_1=m_2=m$ ($\gamma=1$), the effective coupling constant of the phonon and the gapped modes~(\ref{alpha}) reduces to $\alpha_t  \propto (\lambda_{11}-\lambda_{22})/(\lambda_{11} \lambda_{22}-\lambda_{12}^2)$, where the definitions of $\lambda$'s are listed in (\ref{lambdadef}). The phonons are prepared to comprise a wave packet propagating toward the detector with central momentum $k_0$ and width $\Delta k_0$ in the condition of $k_0 +\Delta k_0 < K$ where $K^2=c_{\phi}^2/\zeta_{\phi}=(k^2/2m)^2$  is given by the nonlinear part of the phononic dispersion relation in~(\ref{cs}) in the case of $m_1=m_2=m$ and $g_{11}=g_{22}$. If so, all momentum modes of the sound waves experience approximately the same sound speed $c_{\phi}$ and have common fluctuations in the TOF.
  As a result, the relatively large TOF fluctuations might occur for binary BECs with two hyperfine states  when $ \vert z \vert \rightarrow 1$. Nevertheless, from the mean-field solutions in (\ref{z_a}) and (\ref{z_A}), $ z \rightarrow \pm 1$ is found
in the case of $A/\Omega \gg 1$.   Recall that $A\equiv ( 2 g_{12}-g_{11}-g_{22}) \rho_0/2$ and $\Delta \equiv [(g_{11}-g_{22} ) \rho_0- 2 \delta]/2$ where  one can
tune $a_{12}$ by the Feshbach resonance effect~\cite{kem} and/or $\delta$ by the Zeeman effect, with which  to change $A$ and $\Delta$.  {Another tunable parameter is the Rabi frequency~\cite{zib}. Choice of the small $\Omega$ may lead to large TOF fluctuations. A similar strategy has also been applied by the authors of Ref.~\cite{zib} where the decrease in  the Rabi frequency can enhance the internal Josephson effect given by the ratio $A/\Omega$ in a rubidium
spinor Bose-Einstein condensate, for almost  identical background scattering lengths. Finally, the timescale for switching on and off the interaction $\tau$ is also tunable. The smaller $\tau$ corresponding to a faster switching process is also expected to give the  larger TOF fluctuations to be discussed later.}
 \begin{figure}[b]
\begin{center}
\includegraphics[width=1\linewidth]{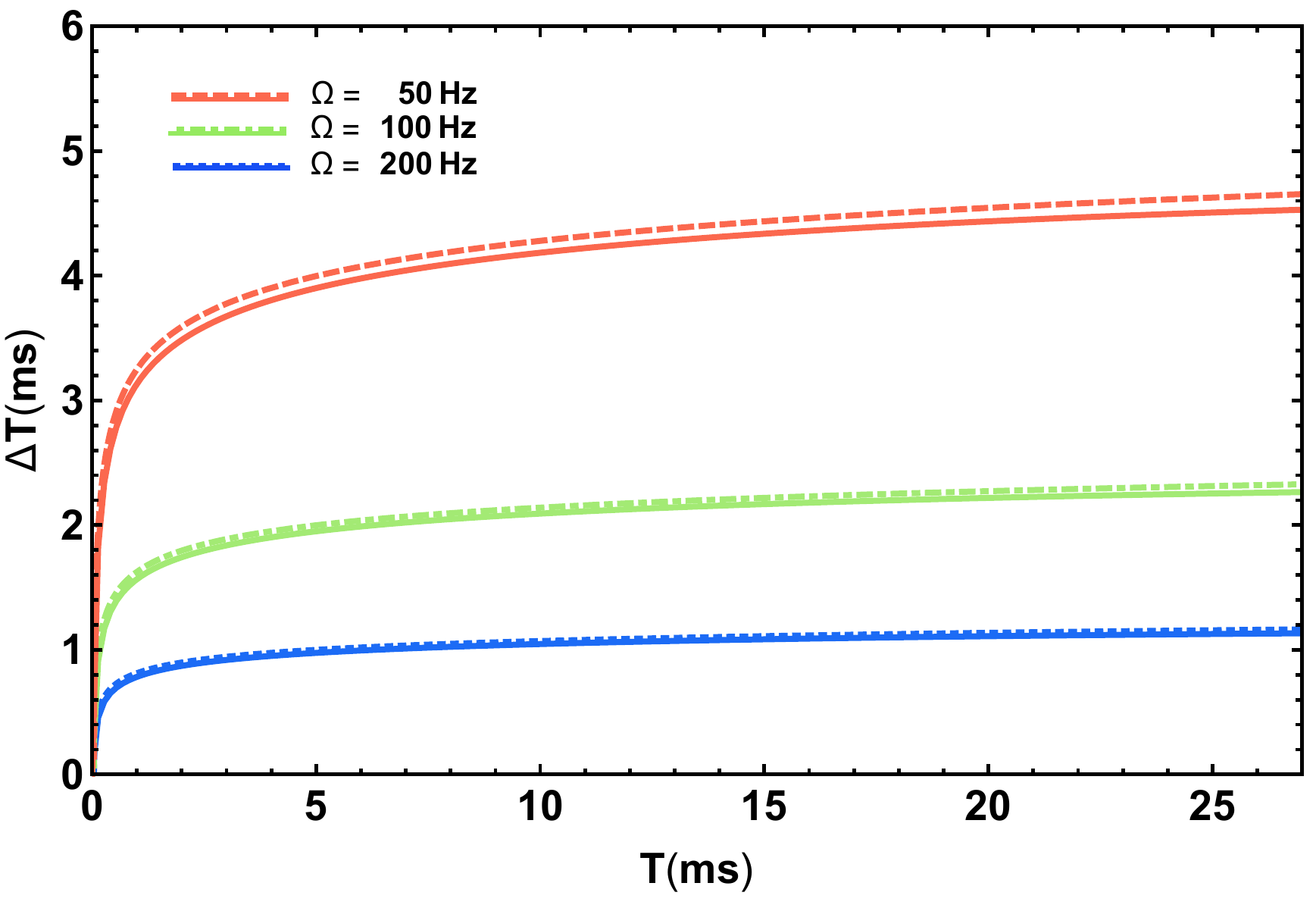}
\caption{The TOF variation $\Delta T $ as a function of $T$ with the scattering length $a_{11}=a_{22}=100\text{a}_0$, $a_{12}=300\text{a}_0$, and detuning $\Delta=0$ with difference choices of Rabi frequency  $\Omega_0$: $\Omega_0=200\text{Hz}$ (blue), $\Omega_0=100\text{Hz}$ (green) and $\Omega_0=50\text{Hz}$ (red).}
\label{DT2_n4}
\end{center}
\end{figure}
The plots in Figs.\,\ref{DT2_n2}(a)--\ref{DT2_n2}(c) show the TOF variation $\Delta T$
as a function of the flight time $T$ with the above-stated parameters, but choosing different $a_{12}$.   In Fig.\,\ref{DT2_n2}(b), the scattering length of  $a_{12}$ is chosen as $a_{12}=300 a_0=15.9 \text{nm}$ where $A/\Omega_0=48.6,\,\, \Delta/\Omega_0=0.0$ with $\Omega_0=200 {\rm Hz}$. The  property of the ground state can be checked from Fig.\,\ref{mean_field_H_A}(e) by choosing $\Omega=24 \Omega_0$ with $z\rightarrow -1 (\rho_{20} \gg \rho_{10})$,  and it is also seen in Fig.\,\ref{mean_field_z_A}(b).
In this case, the speeds of the sound and the gapped modes are equal, $c_{\phi}=c_{\chi}(1.83\mu \text{m}/\text{ms})$~\cite{ham,and} (see Figs.\,\ref{mean_field_summary} and \,\ref{c2} with $\Omega=24 \Omega_0$). According to the experiment in~\cite{and}, the  flight time for the phonon, propagating from the center of the condensates in the axial direction to the distance, say $r=L_z/2 \approx 50 \mu \text{m}$, is about $T \approx 27 \text{ms}$. In Fig.\,\ref{DT2_n2}(b) the flight time fluctuations
grow logarithmic in $T$ to about $
\Delta T \approx 1.12 \,\text{ms}$ with the input timescale of the switching process $\tau=10^{-3}T \approx 0.03 {\rm ms}$ in agreement with the analytical analysis in
Eq.\,(\ref{DT2_t1}). The typical value of $\tau$ is chosen to be  much larger than the timescale given by the inverse of the transition frequency between hyperfine states of the order of radio wave frequencies about ${\rm MHz}$ to ${\rm GHz}$, and it is much smaller than the flight time in this system.
Unfortunately, the effect is not yet testable since
experimentalists most easily measure the (saturated) fluctuations
$\Delta r = c_\phi \Delta T\approx 2.10 \mu \text{m}$  in the position of the
propagating wavefront. Currently, such uncertainty is too small
 with $\Delta r/r \approx 4\%$.

 Choosing $a_{12}=280 a_0=14.84 \text{nm}$ and $a_{12}=320 a_0=16.96\text{nm}$ with the sets of  parameters ($A/\Omega_0=43.7,\; \Delta/\Omega_0=0.0$) and ($A/\Omega_0 =53.5,\; \Delta/\Omega_0=0.0$), giving $c_{\phi} (1.83\mu \text{m/ms}) > c_{\chi} (1.74\mu \text{m/ms})$ and $c_{\phi}(1.83\mu\text{m/ms})  < c_{\chi} (1.92\mu \text{m/ms}) $, in Figs.\,\ref{DT2_n2}(a) and \ref{DT2_n2}(c) we show the respective  flight time fluctuations  as a function of $T$ with their saturated values for large $T$, consistent with the analytic results in~(\ref{DT2_t2}) and (\ref{DT2_t3}). All results of the uncertainty are within $\Delta r/r \approx 4\%$,  the effect is hardly to be measured.

{In what follows, we tune the parameters respectively, including  the scattering length $a_{12}$, the detuning parameter $\delta$,  Rabi frequency $\Omega$, and the timescale $\tau$, to see their effects on the enhancement of the TOF fluctuations.
We first consider to change the scattering length $a_{12}$  from $a_{12}=300 {\rm a_0}=15.9{\rm nm}$ to the value $a_{12}=1000 a_0=53{\rm nm}$ or the detuning parameter $\delta$   from zero up to $100 \Omega_0$, while keeping the other parameters the same as above.
It is found that their tunings do not effectively improve the TOF fluctuation effect, which is still within $\Delta r/r \approx 4{\text {--}}5 \%$,  too small to be seen.
Tuning the timescale $\tau$ of the switching process does not give much improvement on $\Delta r/r$ either.
Nevertheless, reducing the Rabi frequency   is the most efficient way
to give a relatively large  $\Delta r/r $.
Analytically,  as seen in~(\ref{z_A}), setting $\Delta \rightarrow 0$ and $\Omega \ll A$, the mean-field solutions are $z\rightarrow  \pm 1$.   For example,  the solution of $z\rightarrow -1$ leads to $\rho_{20} \rightarrow 1$ but $\rho_{10} \propto \Omega^2$, which is small for small $\Omega$.
However, according to~(\ref{lambdadef}), $\lambda_{11} \propto 1/\Omega^2 , \lambda_{22}, \lambda_{12} \propto {\cal{O}} ({\Omega^0}) $  so as to lead to the TOF fluctuations in~(\ref{ft2}), and $ (\Delta T)^2 \propto 1/\Lambda_{\chi} \propto \lambda_{11} \propto 1/\Omega^2$ with $\Lambda_{\chi}$ in~(\ref{coeffs}). Thus,  the TOF fluctuations are
 inversely proportional to $\Omega^2$ with large effects by reducing Rabi frequency $\Omega$.
Similar arguments are also applied to  the solution of $z\rightarrow +1$.
In Fig.\,\ref{DT2_n4}, the choice of the Rabi frequency is from $\Omega_0=200 {\rm Hz}$ down to $\Omega_0=50 {\rm Hz}$, and the TOF fluctuations can be as large as $\Delta r/r=15 \% $ to be potentially measurable. }

\section{Conclusions}\label{sec5}
The main goal of this work is to introduce the binary BEC system, which is a tunable system  via a Feshbach resonance, with the condensates in different hyperfine states to establish parallels with
stochastic gravity phenomena. Apart from tunable scattering lengths, the Rabi interaction that couples one hyperfine state to the other, and the laser-field induced shift of the energy level of the hyperfine states  characterized by  the detuning parameter $\delta$ are also considered.
We first investigate the properties of homogeneous condensates by a general choice of the tunable interaction parameters, and we find the ground state of the system in terms of the population imbalance $z$, where  $z>0$ corresponds to component 1 dominated and $z<0$ corresponds to component 2 dominated. In particular, when the interaction strength of the different hyperfine states is smaller (larger) than the average of interaction
strengths within each hyperfine state $2g_{12} < g_{11}+g_{22}\,(2g_{12} > g_{11}+g_{22})$, the population imbalance $z$ changes continuously (discontinuously) from $z>0$  to $z<0$
by varying the detuning parameter $\delta$.
These types of phase transitions might
provide a straightforward strategy for establishing parallels
with  the dynamics of the (continuous/discontinuous) phase transitions of the early
Universe on the emergent spacetime given by the time- and/or space-dependent condensates.
We have also studied the perturbations of the system in its energy minimum state by explicitly constructing the Bogoliubov transformation, with which to
identify the corresponding quasiparticle modes.
We show  that the system can be represented by a coupled two-field model of a gapless
 Goldstone phonon and a gapped Higgs mode.
Here we have gone one step further to trace out the gapped modes to give an effective purely phononic theory using closed-time-path formalism. The effects of the gapped modes on the phonons can be calculated explicitly and are all encoded in the influence functional.
As an application of the theory, we are particularly interested in the sound cone fluctuations  due to the possible variation of the speed-of-sound Lorentzian acoustic metric, which are induced by quantum fluctuations of the Higgs gapped modes. To do so, the Langevin equation of the phonon modes under the semiclassical approximation that governs the dynamics of sound waves is derived with the noise term where the correlation function is  given by the Hadamard function of the gapped modes.
Thus, the quantum fluctuations of the gapped modes provide the stochastic component to the sound cone, giving sound cone fluctuations, which are analogous to light cone fluctuations arising from quantum gravity effects.
The effects of sound cone fluctuations can be tested experimentally in principle from the flight time variation in the sound waves.

We have analyzed the resulting expression of the flight time variation as a function of the flight time $T$. It is found that
the integral that sums up the contributions from  all momenta of the gapped modes  has UV divergence for spatial dimension $D\ge 1$. Here we
introduce a suitable switching function, giving a smooth switching process on the Rabi coupling, to
regularize the momentum integral.  We then   focus on the pseudo-one-dimensional problem, which ends up with the UV-finite expression of the flight time fluctuations. In the large flight time limit, the TOF fluctuations can be studied both analytically and numerically  when the sound speed of the phonons is either larger or smaller than that of the gapped modes, and also when  two speeds are the same.
In the case of equal speed, the TOF fluctuations grow logarithmically in $T$ whereas for the other two cases of $c_{\phi}> c_{\chi}$ or $c_{\phi} < c_{\chi}$, the fluctuations are saturated in the large $T$ limit. All analytical results allow us to study their respective effects by changing the tunable parameters.
Given all tunable parameters such as scattering lengths of atoms, the detuning parameter for the shift of the energy levels of the hyperfine states, the Rabi frequency, and the timescale for switching processes,
we conclude  that the most efficient way to
enhance the flight time fluctuations is to reduce the Rabi frequency so as to reach  the fluctuation effect as large as $\Delta T/T \approx 15 \% $, which allows the possibility of experimental confirmation.
The effect of quantum fluctuations looks significant but
whether the effect is measurable, given the other experimental
uncertainties, is unclear at the present time. But its effect is certainly   much larger than numerical estimates for the relative effects of
light-cone fluctuations on TOF of order ${\mathcal{O}}(10^{-9})$, to be
 barely measurable~\cite{ford}.

\begin{acknowledgements}
This work was supported in part by the
Ministry of Science and Technology, Taiwan.
\end{acknowledgements}

\end{document}